\documentclass[preprint,12pt]{elsarticle}

\usepackage{amssymb}

\usepackage{float}
\usepackage{subcaption}
 \usepackage{amsmath}

\journal{}

\begin{document}

\begin{frontmatter}



\title{Pore-scale analysis of gas injection in gas-condensate reservoirs}


\author[inst1]{Paula Kozlowski Pitombeira Reis}

\affiliation[inst1]{organization={Department of Mechanical Engineering, Pontifícia Universidade Católica do Rio de Janeiro},
            addressline={Rua Marques de Sao Vicente, 225}, 
            city={Rio de Janeiro},
            postcode={22451-900}, 
            state={RJ},
            country={Brazil}}

\author[inst1]{Marcio da Silveira Carvalho}

\begin{abstract}
Condensate banking around wellbores can significantly shorten the production from gas-condensate reservoirs. Different approaches to mitigate this issue have been proposed in the literature, among which gas injection comes out with promising results. With this method, pressure maintenance and condensate re-vaporization can be achieved, lessening the flow blockage caused by liquid dropout and accumulation in the porous medium. While gas injection in gas-condensate reservoirs has been largely investigated at the meso and macro scales, data regarding the method's efficiency at the micro scale are scarce. Therefore, the effects of local changes in gas and condensate properties stemmed from the interaction between injected and reservoir fluids at the pore-scale are poorly understood. In order to evaluate how these changes affect the transport in porous media, a compositional pore-network model was used to reproduce gas injection in a sandstone sample following condensate accumulation. $C_1$, $C_2$, $CO_2$, $N_2$ and produced gas were tested as the candidates for condensate banking remediation at different pressure levels. After gas flooding, condensate saturation, heavy component recovery and gas relative permeability were quantified to appraise the achieved gas flow improvement. The results indicated that $C_2$ and $CO_2$ were the most effective gases to clear the accumulated condensate and re-establish the gas flow. Conversely, $C_1$ and $N_2$, especially mixed with the produced fluids, displayed the least favorable results, and could even lead to gas flow impairment.
\end{abstract}




\end{frontmatter}


\section{\label{sec:level1}Introduction}


Production from gas-condensate reservoirs can be significantly affected by condensate banking. As the pressure around producing wells is lowered below the reservoir mixture's dew point, a liquid phase emerges in the porous medium and partially blocks the gas flow. Several parameters affect the severity of the damage, such as the reservoir permeability, fluid composition and depletion level. Even lean gas mixtures can lead to significant flow blockage once liquid dropout takes place \citep{afidick1994production}. The liquid and gas phases different mobilities and molar contents induce a compositional shift in the two-phase region, which enlarges the local heavier component fractions and can lead to substantial liquid saturations \citep{vo2015experimental}.

In order to manage this production challenge, various production strategies and EOR methods have been devised for gas-condensate reservoirs development \citep{sayed2016mitigation,ganie2019review,hassan2019gas}. Among these methods, many involve the injection of gases in the reservoir, aiming full or partial pressure maintenance, as well as re-evaporation of accumulated condensate.

In full pressure maintenance, the reservoir pressure is kept above the dew point pressure and liquid dropout is prevented. If feasible, this method can maximize condensate recovery, as no liquid phase is deposited in the porous medium and heavy components are produced in the gas. \citet{luo2001experimental} conducted long-core flooding experiments with lean gas injection to compare recoveries above and below the dew point pressure. While injection below the dew point managed to re-vaporize a fraction of both intermediate and heavy hydrocarbons in the core, condensate recovery was considerably higher for the case above the dew point. Implementing a full pressure maintenance scheme, however, requires large amounts of injected gas and can be unpractical for cases in which the initial reservoir pressure is close to the fluids dew point \citep{luo2001experimental}. Also, special attention has to be directed to the injected gas composition. Certain gases, such as nitrogen\citep{luo2001experimental} and methane \citep{ahmed1998wellbore}, have a tendency to increase gas-condensate mixtures dew point and can lead to early condensation in the reservoir.

For the case of partial pressure maintenance, gas can be injected following liquid dropout in the reservoir. In this approach, the objectives are both slowing down pressure depletion and re-vaporizing the accumulated condensate, so that gas flowing paths are cleared and valuable heavy components are recovered. Numerous studies have investigated the ability of different gas compositions to support reservoir pressure and reduce liquid banking. \citet{al2004revaporization} evaluated the effectiveness of methane to recover condensate using coreflooding experiments. They concluded that methane is able to re-vaporize the liquid phase and restore the core's original permeability, but that tens to hundreds of injected pore volumes can be required for the method's success. High injection pressures and flow rates were pointed out as means to accelerate the method's outcomes. \citet{gachuz2011laboratory} performed core flooding experiments to investigate the recovery obtained with gas injection in naturally fractured gas-condensate reservoirs. A small gap was allowed around the core to simulate a fracture and, after the core was depleted below the dew point pressure, $N_2$, $CO_2$ and lean gas were injected. The results indicated a condensate recovery of $51.7\%$ with the lean gas injection, $34.78\%$ with $CO_2$ and only $18.7\%$ with the $N_2$ injection case. \citet{al2012mobility} evaluated condensate recovery by the injection of super-critical $CO_2$, $C_1$, and their mixtures. They performed gas injection in a sandstone core following condensate flooding, meaning that the liquid content was not established in the porous medium by condensation. Unsteady-state
relative permeability and recovery measurements demonstrated that the injection of pure $CO_2$ delayed the gas breakthrough and provided higher recovery of condensate when compared to the injection of mixtures of $CO_2$ and pure $C_1$, at the same operating conditions. Besides conventional gas flooding, huff-n-puff gas injection has also been investigated experimentally, for application in shale gas-condensate reservoirs. Due to shale ultra low permeabilities, this technique could be beneficial to expedite both pressure boost and re-evaporation of condensate in the producing wellbore vicinity \citep{sheng2015increase}. Core flooding experiments in shale cores displayed promising results with the huff-n-puff injection of produced gas \citep{meng2016experimental}, $C_1$ \citep{sharma2018comparative} and $CO_2$ \citep{meng2018performance}.

Additionally to core flooding tests, several reservoir-scale numerical studies have been conducted to assess gas injection as an enhanced gas-condensate recovery method. \citet{marokane2002applicability} used a full-field compositional reservoir simulation model to investigate the efficacy of a one-time lean gas injection scheme to remove the condensate banking around producing wells. Results indicated that, for a lean gas reservoir, maximum recovery was obtained for injection at pressures below the maximum liquid dropout. For a rich gas reservoir, however, starting gas injection before reaching the maximum liquid dropout pressure produced better results. \citet{linderman2008feasibility} used a similar approach to evaluate the feasibility of injecting nitrogen in a gas-condensate reservoir. Their model predicted superior net gas production with the injection of nitrogen than with the injection of produced gas, while the condensate net recovery was virtually identical for the two scenarios. They also compared the injection of $CO_2$ with $N_2$, finding no significant differences in the achieved recoveries. \citet{taheri2013miscible} modeled enhanced condensate recovery in fractured gas-condensate reservoirs using stock tank gas, $C_1$, $N_2$ and $CO_2$ injection. Results suggested that $CO_2$ injection could generate considerably better results that the other tested gases, and that no appreciable difference is obtained between the injections of methane, nitrogen and stock-tank gas. It was also concluded that injection at higher pressures boosted condensate recovery. Another numerical study evaluating gas injection in fractured gas-condensate reservoirs was conducted by  \citet{fath2016investigation}. After testing four different injected gas compositions, they obtained the highest recovery with $CO_2$, followed by $C_1$, produced gas and $N_2$. Their model also predicted that, for the same injected volumes of the tested gases, nitrogen led to a significant higher pressure average in the reservoir, while produced gas led to the lowest. Recent numerical works also investigated the efficacy of $C_2$ injection for condensate recovery, with very favorable results. \citet{sharma2017comparative} evaluated the performance of huff-n-puff gas and solvent injection in shale gas-condensate reservoirs. They concluded that ethane was the best injection fluid on accounts of higher and much faster recovery, when compared to methane, methanol and isopropanol. The positive results were attributed to ethane's ability to greatly reduce the gas-condensate mixture's dew point pressure, ensuring relatively low injecting volumes for condensate re-vaporization. \citet{zhang2020investigation} investigated the injection of $C_1$, $C_2$, $CO_2$ and $N_2$ in a five-spot configuration for condensate enhanced recovery. In their study, $C_2$ also led to the best condensate recovery, of $56.8\%$ after the injection of 0.85 pore volumes, closely followed by $CO_2$, with a recovery of $50.7\%$. Methane and nitrogen displayed considerably worse performances, with $22.8\%$ and $21.3\%$ of condensate recovery, respectively, for the same injected volume.

Although both experimental and numerical research indicate that gas injection is an advantageous approach for condensate enhanced recovery, no micro-scale analysis of the method has been reported in the literature. This represents a significant gap in data for gas injection performance evaluation, especially considering that pore-scale events are essential to understanding macro-scale transport properties in porous media. During gas injection, local changes in composition can alter significantly both bulk and interfacial properties of gas and liquid phases, which in turn affect their flow, characterized by the relative permeability curves. To address this gap, we used a compositional pore-network model to evaluate gas injection in porous media after condensate accumulation. The model has been validated against core flooding experiments \citep{reis2021pore} and also used to evaluate wettability alteration as a gas-condensate enhanced recovery method \citep{reis2020pore}. In the present study, first we replicated condensate accumulation in the porous medium by flowing a representative gas-condensate mixture through a sandstone-based network at different depletion levels, starting from values just below the dew point, until pressures below the maximum liquid dropout. Then $C_1$, $C_2$, $CO_2$, $N_2$ or produced gas were injected in the condensate bearing network and the flow improvement was evaluated. Final saturations, relative permeabilities and recovery of heavy hydrocarbon components were quantified to compare the efficacy of each injection scenario. The impact of the concentration of the injected gases was also explored, by repeating the analysis with the injection of mixtures containing $50\%$ in moles of $C_1$, $C_2$, $CO_2$ or $N_2$, and $50\%$ the produced gas.

\section{Pore-network model}

The pore-network model used in the preset study was devised specifically to represent gas-condensate flow in porous media. It was based on the fully compositional pore-network model proposed by \citet{santos2020pore} and adapted to encompass realistic characteristics of \textit{in-situ} condensate formation and gas-condensate flow. In the following sections, the main aspects of gas-condensate flow in porous media and their representation in the model are described. For further details, we refer to \citet{reis2021pore}.

\subsection{Pore-scale gas-condensate flow patterns}
\label{sec:gc_flow}
Gas-condensate mixtures undergo a transition from single-phase gas flow to two-phase gas-condensate flow, as the dew point pressure is reached. Upon condensation,
it has been observed in micro model experiments \citep{cocskuner1997microvisual,al2009condensate} that the liquid phase wets completely the porous medium and tends to form a film that flows adjacent to the walls, while the gas flows along the pore centers, in a core-annular pattern. This flow configuration can be disrupted, however, if the film becomes unstable and the snap-off of the gas happens. In this case, a bridge of liquid is formed across the gas flowing channel and blocks the flow. Once blocked, a balance between viscous and capillary forces determines whether the channel is re-opened to gas flow. This intermittent opening and closing of gas flowing channels  is pointed out as the main mechanism behind condensate blockage \citep{jamiolahmady2003positive}.

\begin{figure}[H]
    \centering
    \includegraphics[width=40mm]{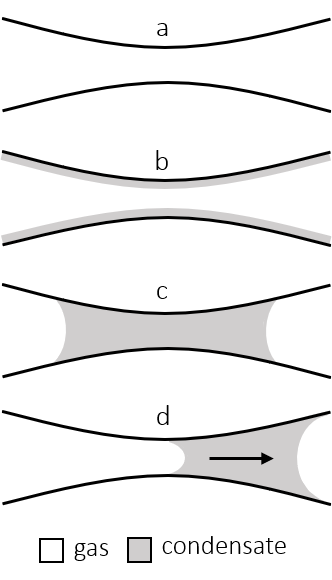}
    \caption{Evolution of condensate configuration in a pore throat \cite{reis2021pore}}
    \label{fig:cond_ev}
\end{figure}

In our model, we represented this mechanism as illustrated in Figure \ref{fig:cond_ev}. Figure \ref{fig:cond_ev}(a) represents a flowing channel above the dew point pressure, where only gas flows. Figure \ref{fig:cond_ev}(b) depicts the flowing channel below the dew point pressure. In this case, the phases flow in core-annular configuration, with condensate at the walls and gas at the center. Figure \ref{fig:cond_ev}(c) contains the phases configurations once the snap-off of the gas happens and the bridge of liquid is formed. Finally, Figure \ref{fig:cond_ev}(d) illustrates the re-opening of the flowing channel, with the mobilization of the condensate bridge. 

\subsection{Pore Geometry}
\label{pn_geom}

\begin{figure}[H]
    \centering
    \includegraphics[width=70mm]{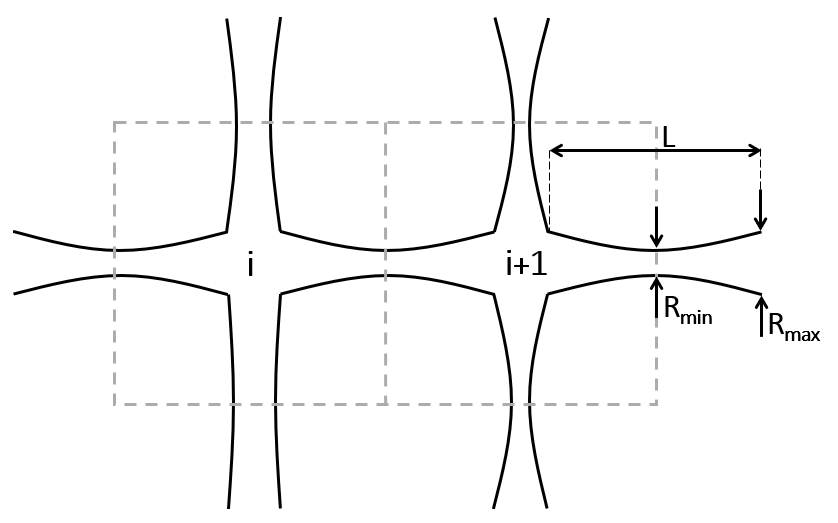}
    \caption{Definition of a pore volume \cite{reis2020pore}}
    \label{fig:volumeporo}
\end{figure}

 Networks of pore bodies connected by converging-diverging circular pore throats were used in the model. Equation \ref{eq:r_fun} expresses the radius of a pore throat $j$, with length $L_j$, maximum radius $R_{max,j}$ and constricted radius $R_{min,j}$. Control volumes for each pore $i$ were defined encompassing half the volumes of the throats connected to them, as illustrated in Figure \ref{fig:volumeporo}. The constricted pore throat geometry was chosen so that an accurate criterion for the snap-off of the gas phase could be implemented. Apart from that, it allows the calculation of unique and continuously varying capillary pressure as the the liquid bridge meniscii move in a pore throat \citep{al2005dynamic} (e.g. Figure \ref{fig:cond_ev}(c) and (d) ). Consequently, 
 the conditions for gas flow blockage could be adequately represented.

 \begin{subequations}
\label{eq:r_fun}
\begin{align}
 r_j(x) = \sqrt{a_j+b_jx^2}  \\
 a_j =  R_{min,j}^2    \\
 b_j =  \left(\frac{2}{L_j}\right)^2(R_{max,j}^2-R_{min,j}^2)
 \end{align}
\end{subequations}

\subsection{Fluid conductances and snap-off criteria}
\label{sec:snap_off}

Regarding the two-phase flow calculations, it was considered that pore bodies impose no restriction to the flow, so that conductances were devised only for pore throats. For the annular flow developed after condensate dropout, the conductances for the gas and liquid phases are presented in equations \ref{eq:g}(a) and \ref{eq:g}(b), respectively.

\begin{subequations}
\label{eq:g}
\begin{align}
 g_{g} &=\frac{\pi}{8\mu _{g} L}(S_{g} R^2)^2+\frac{\pi}{4\mu _{l} L}(R^2-S_{g} R^2) S_{g} R^2  \\
 g_{l} &=\frac{\pi}{8\mu _{l} L}(R^2-S_{g} R^2)^2 
 \end{align}
\end{subequations}

\noindent where, $\mu _g$ and $\mu _{l}$ are the viscosities of the gas and liquid phases, $S_{g}$ is the gas saturation and $R$ and $L$ are, respectively, the equivalent radius\citep{sochi2013newtonian,reis2021pore} and length of the capillary. 

The snap-off of the gas phase takes place in a pore throat once a critical thickness of the liquid film is achieved  \cite{reis2021pore} and the criterion proposed by  \citet{beresnev2009condition} is met. This criterion, given by equation \ref{eq:berev}, takes into consideration the length, maximum and minimum radii of the throat, as well as the condensate film thickness ($e$). 

\begin{equation}
\label{eq:berev}
L>2\pi\sqrt{(R_{min}-e)(R_{max}-e)} 
\end{equation}

For the flow through a pore throat to be reestablished after the snap-off, the pressure drop across the throat has to overcome the critical value given by equation \ref{eq:dP_crit}. It is a function of the interfacial tension, $\sigma$, between the phases and the radii of the condensate bridge meniscii, $R_1$ and $R_2$. The interfacial tension is calculated with the correlation proposed by \citet{WK}, and the radii vary according to the throat geometry and the condensate saturation.

\begin{equation}
\label{eq:dP_crit}
\Delta P_{crit}=2\sigma \left(\frac{1}{R_{1}}-\frac{1}{R_{2}}\right)
\end{equation}

\subsection{Governing Equations}
\label{sec:ge}
A system of non-linear equations relates the molar content and pressure of each pore control volume $i$ in the model. It encompasses molar balance equations (eq. \ref{eq:mb}), volume consistency equations (eq. \ref{eq:vc}), and equations to enforce boundary conditions. 

 Equation \ref{eq:mb}(a) describes how the number of moles in a pore volume $i$ varies due to the molar flow rate through its adjacent pore throats, $\dot{n}^k_j$, and the flow at the network boundaries, $\dot{s}_i^k$. In this equation, $c_{ij}$ are the entries of the incidence matrix $C$, which maps which throats are connected to each pore volume in the network. The molar flow through a pore throat, given by equation \ref{eq:mb}(b), converts the volumetric flow, calculated with the conductances and the pressure drop, into the molar flow, using the molar fraction of each component in the gas and liquid phases, $y^k$ and $x^k$, and their molar densities, $\xi_g$ ans $\xi_l$. Equation \ref{eq:mb}(c) represents the pressure difference that drives the flow through the throats. The inclusion of the interface pressure difference ($\Delta P_j^{int}$) in eq. \ref{eq:mb}(c) is controlled by the parameter $H^{int}$, which is equal to $1$ when a liquid bridge is present and $0$, otherwise. In Equations \ref{eq:mb}, $i=1..n_b$, $j=1..n_t$, and $k=1..n_c$, represent, respectively, the number of pore volumes, pore throats and fluid mixture components in a network.

\begin{subequations}
\label{eq:mb}
    
\begin{equation}
\frac{\partial N^k_i}{\partial t}=-\displaystyle\sum_{j=1}^{n_{t}} c_{ij}\dot{n}_j^k+\dot{s}_i^k
\end{equation}

\begin{equation}
    \dot{n}_j^k =(y^k\xi _{g}g_{g}+x^k\xi _{l}g_{l})_j\Delta P_j \\
\end{equation}

\begin{equation}
\Delta P_j =\displaystyle\sum_{m=1}^{n_{b}} c_{mj}P_m - H_j^{int}\Delta P_j^{int}
\end{equation}

\end{subequations}

The volume consistency equations, given by eq. \ref{eq:vc}, are used to enforce compatibility between the pore volumes and the volumes of the phases contained in them. These equations are devised considering slightly compressible networks, so that, for a given pressure $P_i$, the volume of a pore can be approximated using the pore compressibility, $\nu _i$, along with reference volume, $\overline{V_i}$, and pressure, $\overline{P_i}$ values. For the calculation of the gas and liquid phases volumes, the pressure and temperature in the pore, $P_i$ and $T$, are related with the fluid parameters $\mathcal{L}_i$, $Z_i^{g}$, $Z_i^{l}$, $x_i^k$ and $y_i^k$ (respectively: the fraction of the $N_i$ moles in the liquid phase,  the compressibility factors and the molar fractions of each component $k$ in the gas and liquid phases).

\begin{equation}
    \label{eq:vc}
    N_i-\frac{\overline{V_i}[1+\overline{\nu _i}(P_i-\overline{P})]}{\mathcal{L}_i\left(\frac{Z_i^{l}RT}{P_i}-\sum_{k=1}^{n_c} v_kx_i^k\right)+(1-\mathcal{L}_i)\left(\frac{Z_i^{g}RT}{P_i}-\sum_{k=1}^{n_c} v_ky_i^k\right)}=0
\end{equation}

Additionally to equations \ref{eq:mb} and \ref{eq:vc}, simple equations are written to enforce the boundary conditions for the system. In the presented model, the flow can be driven either by imposing different pressure levels at the inlet and the outlet, or by prescribing molar flow rate at the inlet and pressure at the outlet. Other parameters that have to be imposed are the composition of the fluid injected in the network and the temperature $T$.

The presented system of equations was based on the compositional formulation proposed by \citet{collins1992efficient}, which decouples the flow equations from the flash calculations. In this formulation, the system relating flow equations is solved with the Newton-Raphson method, while thermodynamic equilibrium is enforced at each newton iteration by performing phase equilibrium calculations using the Peng-Robinson EoS and the trial solution variables. Details of the implemented phase equilibrium calculations can be found in \citet{santos2020pore}.

\section{Results}

\subsection{Pore-network}

\begin{figure}[H]
    \centering
        \includegraphics[height=2in]{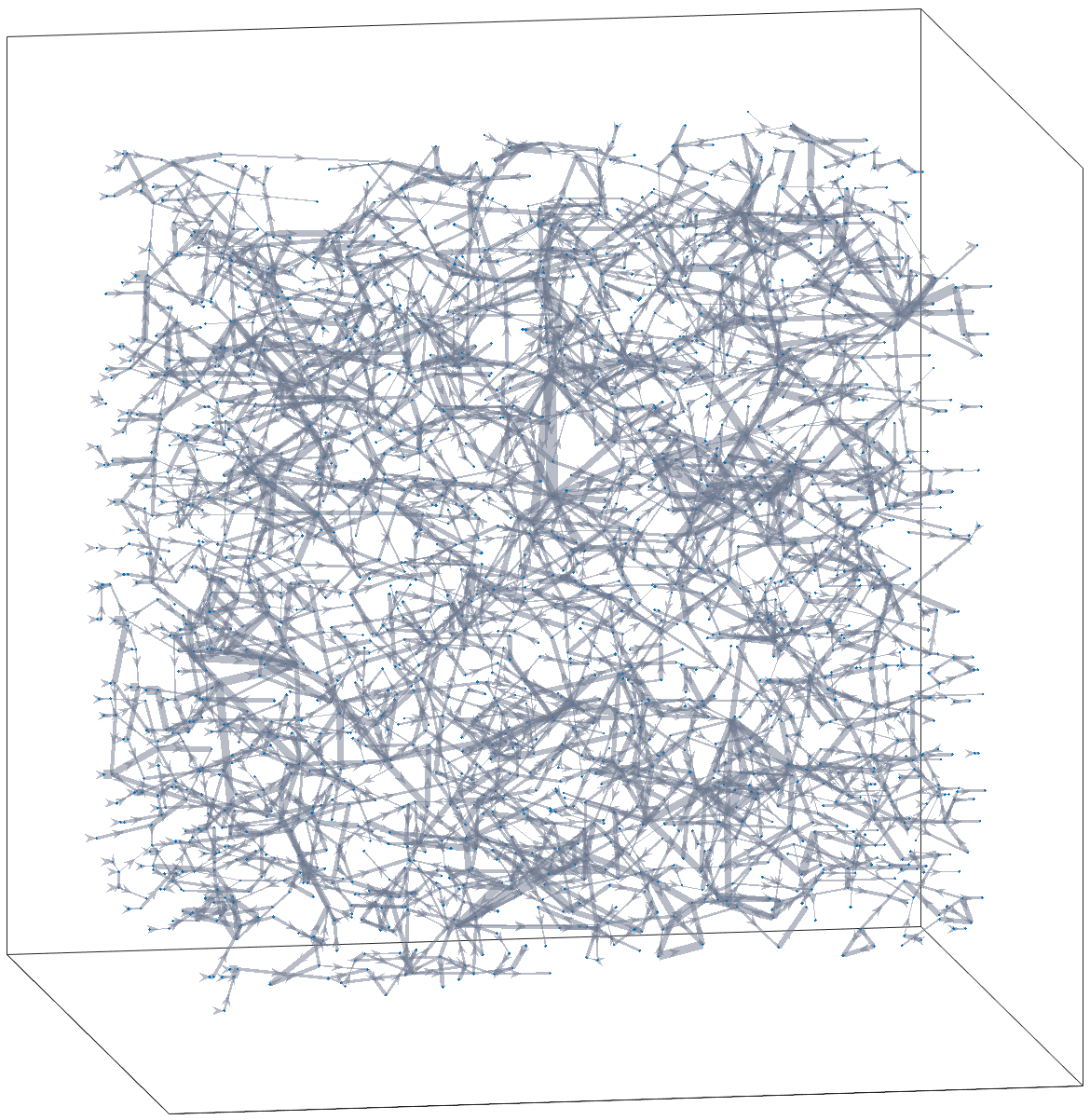}
        \caption{Sandstone based network}
    \label{fig:net}
\end{figure}

\begin{figure}[H]
    \centering
        \includegraphics[height=2.5in]{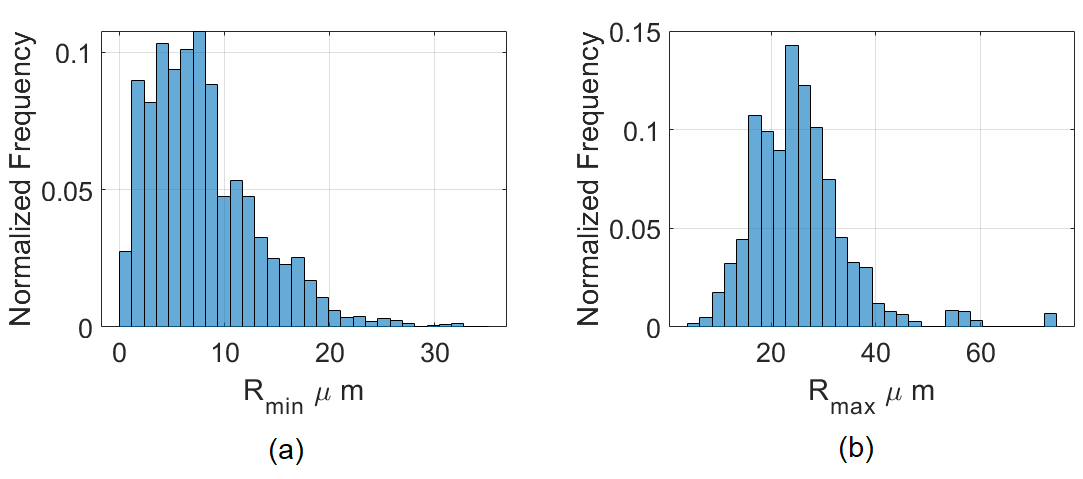}
        \caption{$R_{min}$ and $R_{max}$ distributions in the used pore-network}
    \label{fig:net_hist}
\end{figure}

The pore-network used in the present study, shown in Figure \ref{fig:net}, represented a $1.75 mm^3$ cubic section of a sandstone with permeability of $169mD$ and porosity of $17.1\%$. This network was first extracted from Micro-CT imaging of a sandstone sample, using the network extraction algorithm proposed by \citet{dong2008micro}, then adapted to meet the geometric criteria described in section \ref{pn_geom}. Figure \ref{fig:net_hist} represents the size distributions of the constricted and unconstricted capillary radii of the adapted pore-network. More details of the network adaptation process can be found in \citet{reis2020pore}. 

\subsection{Fluids}

A series of fluid mixtures was used in the present study in order to represent condensate buildup in porous medium followed by gas injection and condensate re-vaporization.

\begin{figure}[H]
    \centering
    \includegraphics[width=100mm]{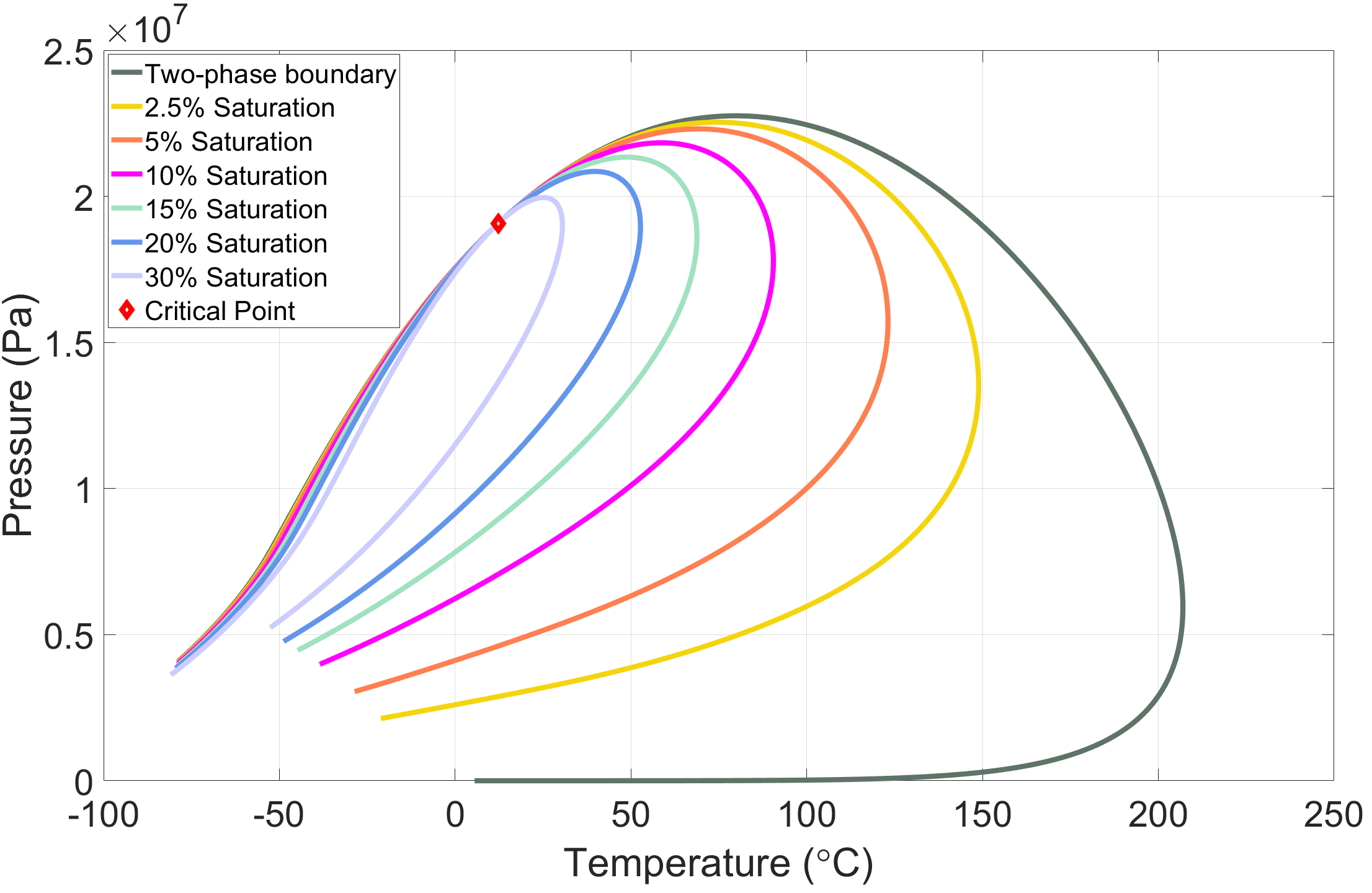}
    \caption{Fluid mixture phase envelope}
    \label{fig:pe}
\end{figure}

The composition of the fluid used during the condensate accumulation process in the porous medium is presented in Table \ref{tab:comp}. This mixture represents a typical fluid found in gas-condensate reservoirs, composed by carbon dioxide, nitrogen, light, intermediate and heavy hydrocarbons. It exhibits a large window of retrograde condensation behavior in the temperature range from $12^{\circ}C$ to $207^{\circ}C$, as shown in the phase diagram in Figure \ref{fig:pe}. For our analyses, a temperature of $60^{\circ}C$ was chosen as the reservoir temperature.

\begin{table}[H]
\centering
\caption{\label{tab:comp}Gas-condensate mixture composition.}
\begin{tabular}{cc}
Component & Molar Fraction\\
\hline
$CO_2$	    & 0.05\\
$N_2$	    & 0.02\\
$C_1$		& 0.65\\
$C_2$		& 0.13\\
$C_3$		& 0.07\\
$C_6$		& 0.05\\
$C_{10}$		& 0.025\\
$C_{16}$		& 0.005\\
\end{tabular}
\end{table}

\begin{figure}[H]
\centering
    \begin{subfigure}[t]{0.5\textwidth}
       \centering
        \includegraphics[width=80mm]{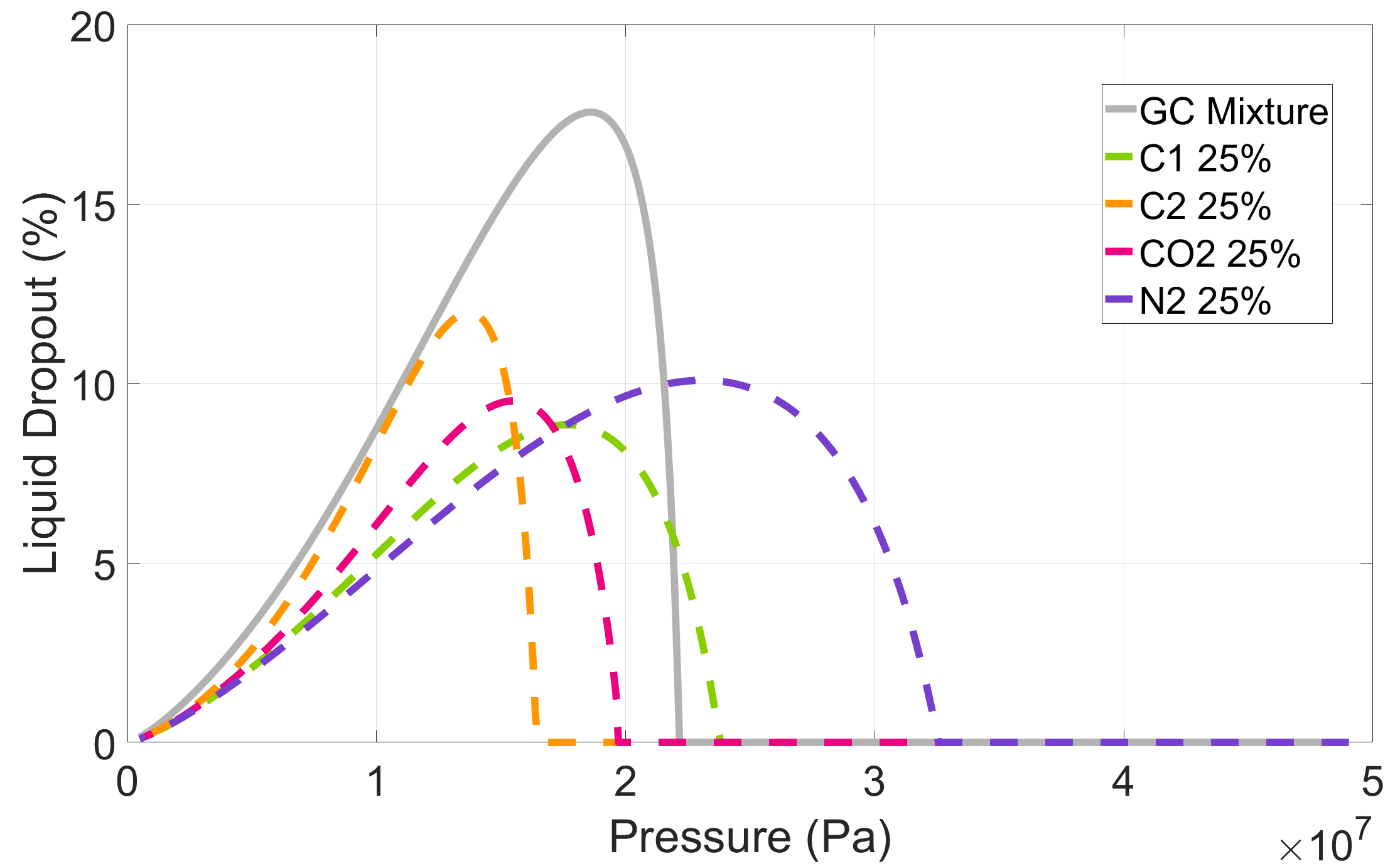}
        \caption{ 75\% of the gas-condensate mixture and 25\% of the injected gases }
    \end{subfigure}%
    
    \begin{subfigure}[t]{0.5\textwidth}
        \centering
        \includegraphics[width=80mm]{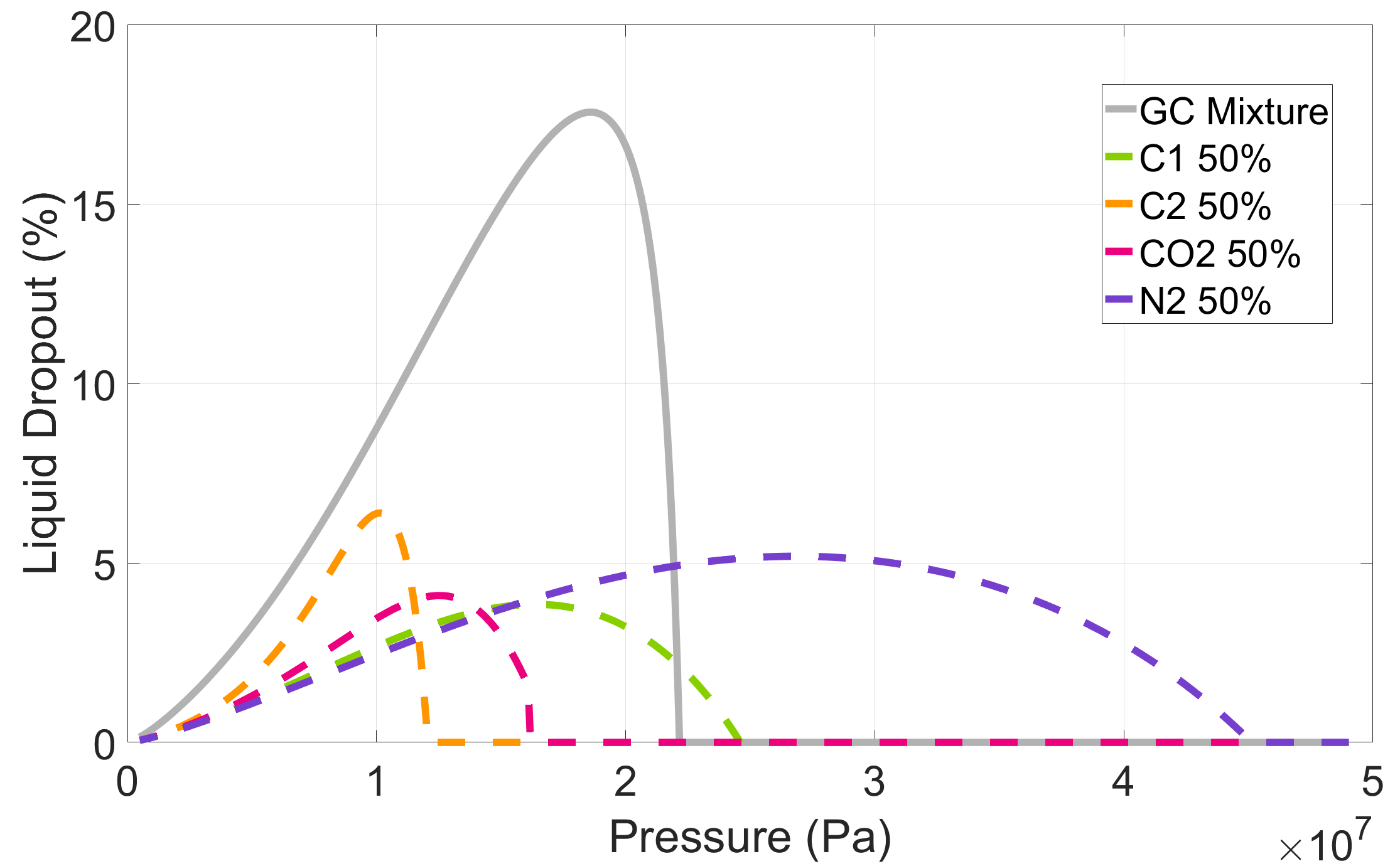}
        \caption{ 50\% of the gas-condensate mixture and 50\% of the injected gases }
    \end{subfigure}
    
    \begin{subfigure}[t]{0.5\textwidth}
        \centering
        \includegraphics[width=80mm]{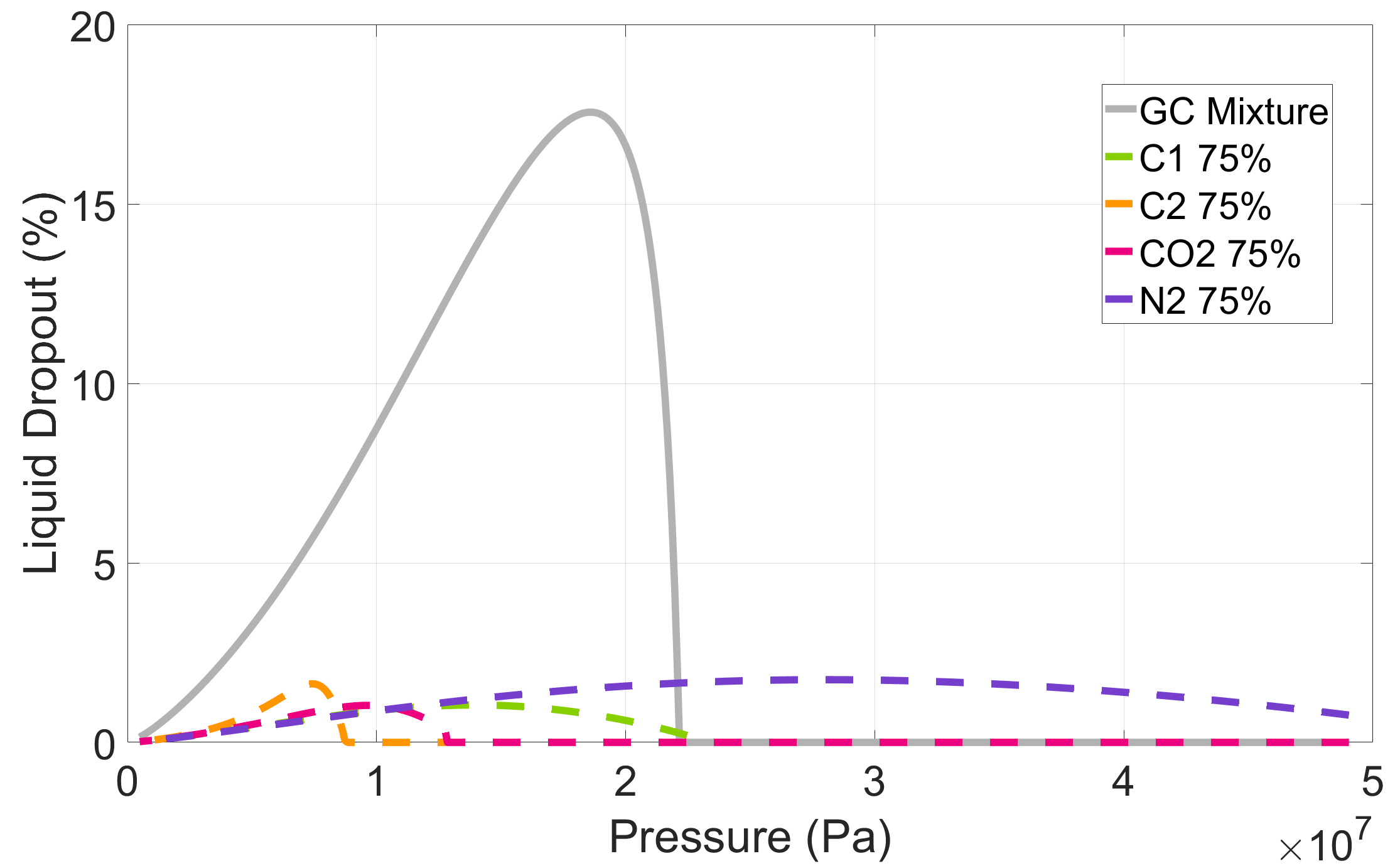}
        \caption{ 25\% of the gas-condensate mixture and 75\% of the injected gases }
    \end{subfigure}
    
    \caption{Liquid dropout at a temperature of $60^{\circ}C$}
    \label{fig:LDO}
\end{figure}

After condensate accumulation, the gases chosen as candidates for injection aiming condensate enhanced recovery were produced gas obtained during the flow of the mixture presented in Table \ref{tab:comp}, methane ($C_1$), ethane ($C_2$), carbon dioxide ($CO_2$) and nitrogen ($N_2$). For the sake of appraising preliminarily the ability of the four last gases to re-vaporize condensate, Figure \ref{fig:LDO} illustrates the liquid dropout as a function of the pressure obtained by mixing $C_1$, $C_2$, $CO_2$ or $N_2$ with the composition presented in Table \ref{tab:comp}, at different molar fractions and a temperature of $60^{\circ}C$.
It can be noticed from Figure \ref{fig:LDO}(a) that the addition of $25\%$ in moles of the tested gases to the reservoir composition leads to a substantial reduction in maximum liquid dropout, varying from $31.8\%$, in the case of $C_2$, to $49.6\%$, in the case of $C_1$. It is also noticeable that the effects of the tested gases on the mixture's dew point pressure are very distinct. $C_2$ and $CO_2$ reduce the saturation pressure, thus allowing the mixture to remain as a single gas phase at lower pressures. The opposite is verified as the reservoir fluid is mixed with $C_1$ and $N_2$. While the negative effect of $C_1$ on the dew point pressure is mild, $N_2$ increases significantly its value, and can lead to very early condensation in porous media. The same trend of effects is observed as the content of the tested gases is increased to $50\%$, Figure \ref{fig:LDO}(b), and $75\%$, Figure \ref{fig:LDO}(c). The maximum liquid dropout is progressively reduced, with $C_1$ and $CO_2$ being the most, and $C_2$ and $N_2$ the least effective gases to lower the maximum volume of liquid formed. As for the effects on the dew point pressure, at higher molar fractions $C_2$ and $CO_2$ gradually reduce it, while $N2$ increases it. $C_1$, on the other hand, slightly augments and then reduces the mixture's dew point pressure, at the molar fractions of $50\%$ and $75\%$, respectively. These results suggest that mixtures rich in $C_1$, $C_2$, $CO_2$ and $N_2$ have a positive prospect to work as condensate re-vaporization agents and, consequently, improve gas flow in porous media.

Additionally to the injection of $C_1$, $C_2$, $CO_2$, $N_2$ or produced gas individually, mixtures of $C_1$, $C_2$, $CO_2$ and $N_2$ with the produced gas, at molar fractions of $50\%$, were also injected in the networks to assess their ability to recover condensate after liquid banking. These tests were devised considering two different scenarios. First, to take advantage of the produced gas availability, we wanted to test whether directly injecting mixtures of the produced gas with $C_1$, $C_2$, $CO_2$ or $N_2$ could satisfactorily enhance condensate recovery. Secondly, even during the injection of pure gases, the flow of mixtures of injected and reservoir fluids will take place, especially at regions far from the injection wells and/or at small injection volumes. These test, therefore, serve also to appraise the effects of in situ mixing with the reservoir fluid in the performance of $C_1$, $C_2$, $CO_2$ and $N_2$ to recover condensate.

\subsection{Flow conditions}

Two steps of injection were performed in the networks for every evaluated case. First, the reservoir fluid was injected so that condensate accumulated in the porous medium, mimicking the process of condensate banking in the near wellbore regions. Then, the injected composition in the networks was altered, so that the liquid buildup was followed by gas injection and condensate recovery.

The process of condensate accumulation was reproduced by injecting 25 pore volumes (PV) of the mixture presented in Table \ref{tab:comp} in the networks at different pressures. As the timing for gas injection is one of the most relevant parameters for the method's success \citep{marokane2002applicability}, we wanted to test its performance at various reservoir depletion stages. Six pressure values were used, 22, 21.5, 21, 19.5, 17.75 and 14.75 MPa, which corresponded to liquid dropout saturations of $2.3\%$, $9.7\%$, $13.3\%$, $17.1\%$, $17.3\%$ and $14.7\%$, respectively. With that, a broad range of depletion scenarios could be covered, from the early stages of condensate formation to pressures below the maximum liquid dropout. 

After the first 25 PV of injection, $C_1$, $C_2$, $CO_2$, $N_2$, produced gas or their mixtures were injected in the the networks. This step also lasted until 25 PV were injected. Under certain conditions, the recovery of heavy components by gas injection can be a very slow process, requiring tens to hundreds of pore volumes injected \citep{al2004revaporization}. For this reason, in some cases, we could not reach a steady state flow. 

The prescribed boundary conditions for both steps were molar flow rate at the network inlet and pressure at the outlet. For all tested injection scenarios, a molar flow rate was imposed so that a gas flowing velocity of $35m/day$ was achieved. It has been reported in the literature \citep{al2004revaporization} that injection rate does not influence significantly the performance of gas injection as a gas-condensate EOR method and we chose, therefore, not to explore the effects of this parameter. 

\subsection{Condensate saturation reduction}
\label{sec:cond_sat}

\begin{figure}[H]
     \begin{minipage}[l]{0.45\textwidth}
         \centering
          \begin{subfigure}{1\textwidth}
            \includegraphics[width=7cm]{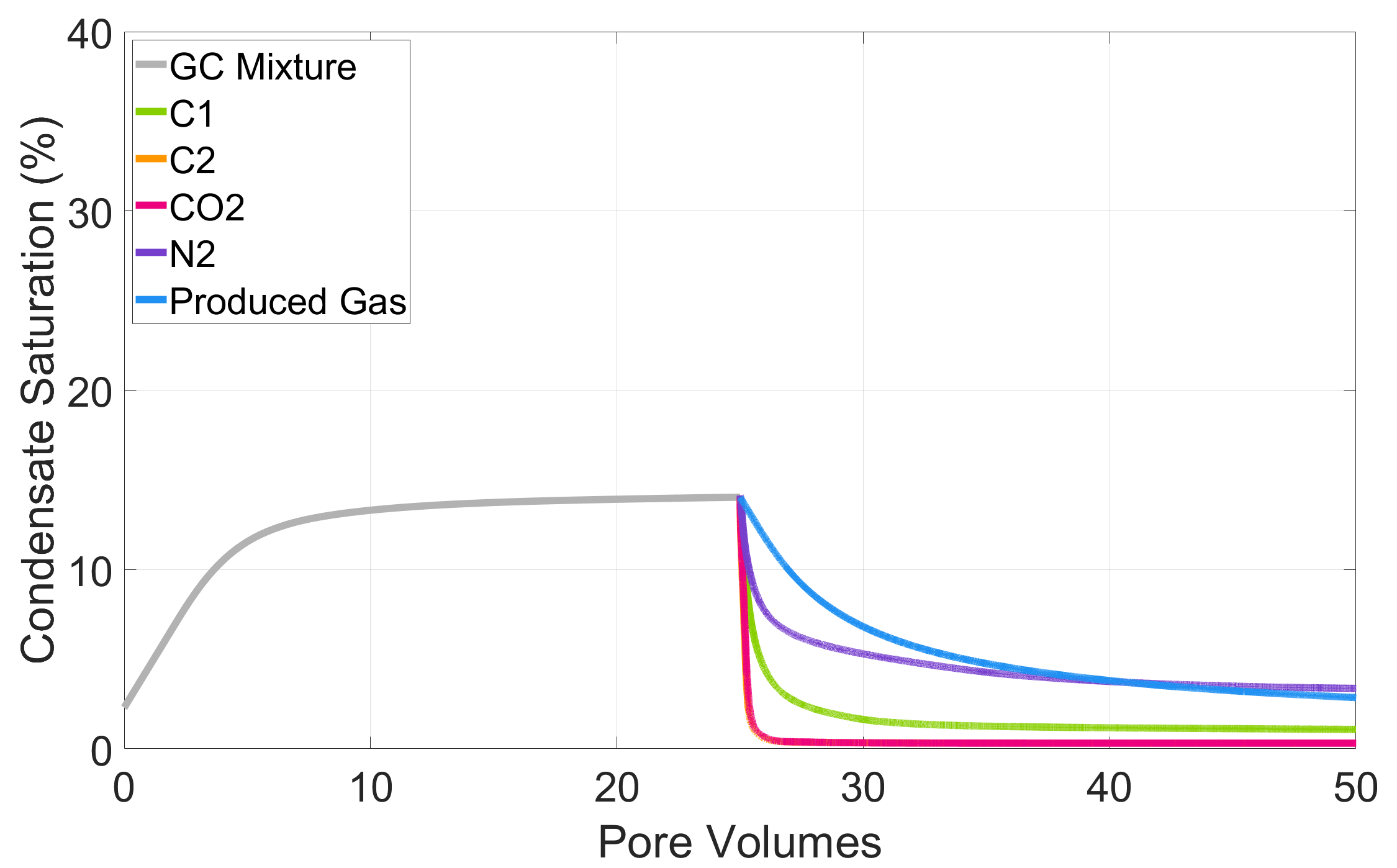}
            \caption{$P=22 MPa$}
         \end{subfigure}
     \end{minipage}
     \hfill{}
     \begin{minipage}[r]{0.45\textwidth}
         \centering
          \begin{subfigure}{1\textwidth}
            \includegraphics[width=7cm]{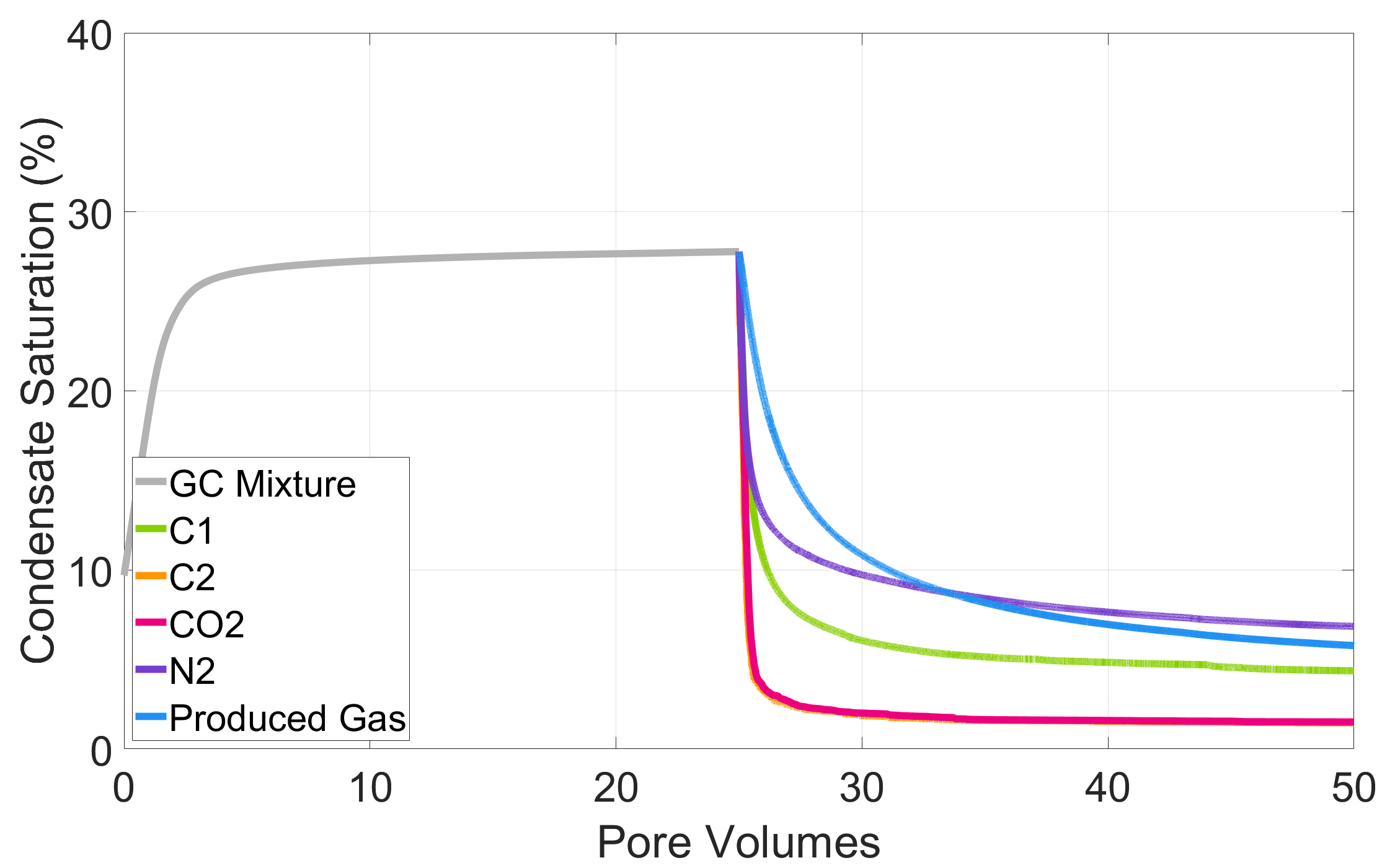}
            \caption{$P=21.5 MPa$}
         \end{subfigure}
     \end{minipage}
     \hfill
     \begin{minipage}[l]{0.45\textwidth}
         \centering
          \begin{subfigure}{1\textwidth}
            \includegraphics[width=7cm]{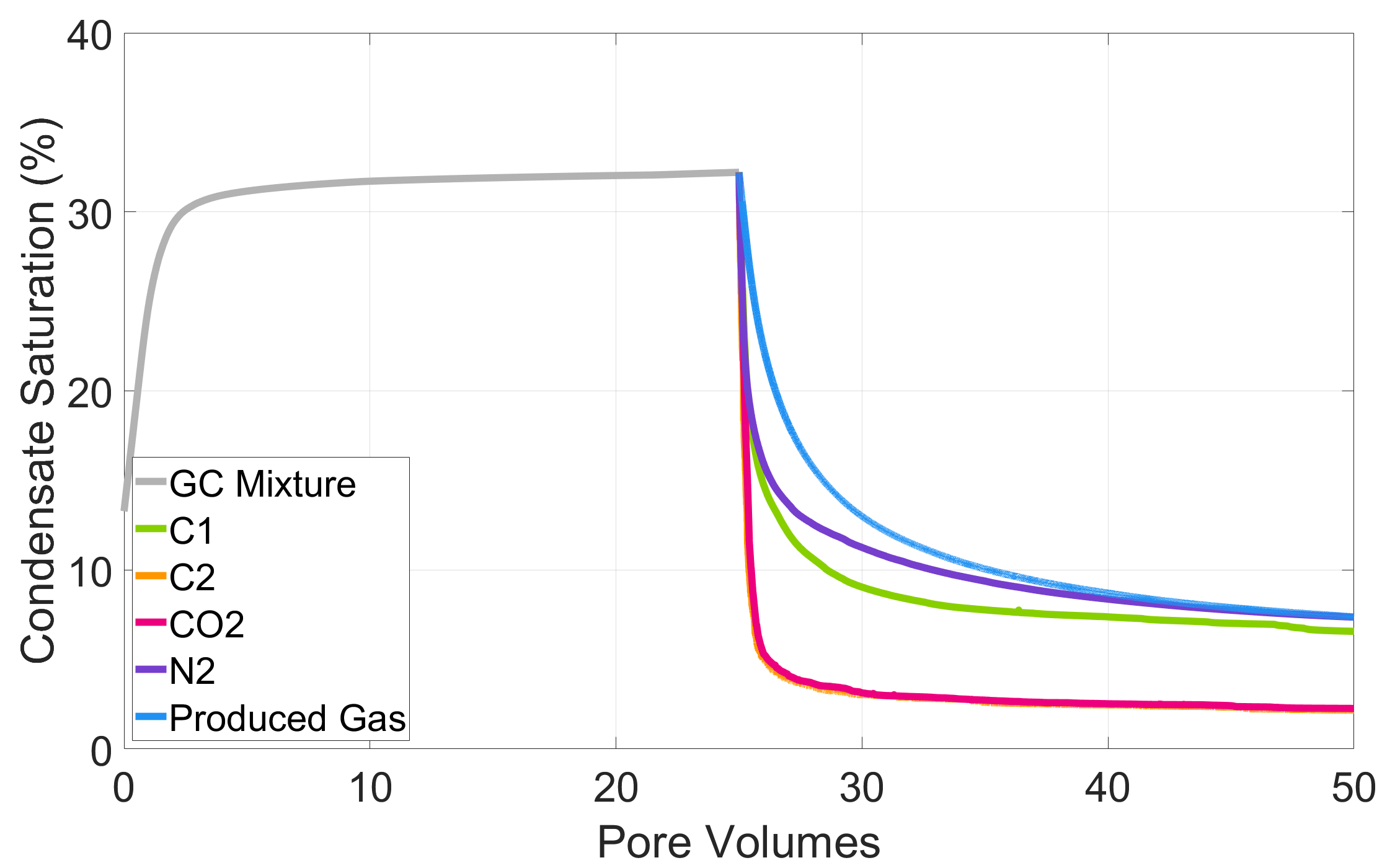}
            \caption{$P=21 MPa$}
         \end{subfigure}
     \end{minipage}
     \hfill{}
     \begin{minipage}[r]{0.45\textwidth}
         \centering
          \begin{subfigure}{1\textwidth}
            \includegraphics[width=7cm]{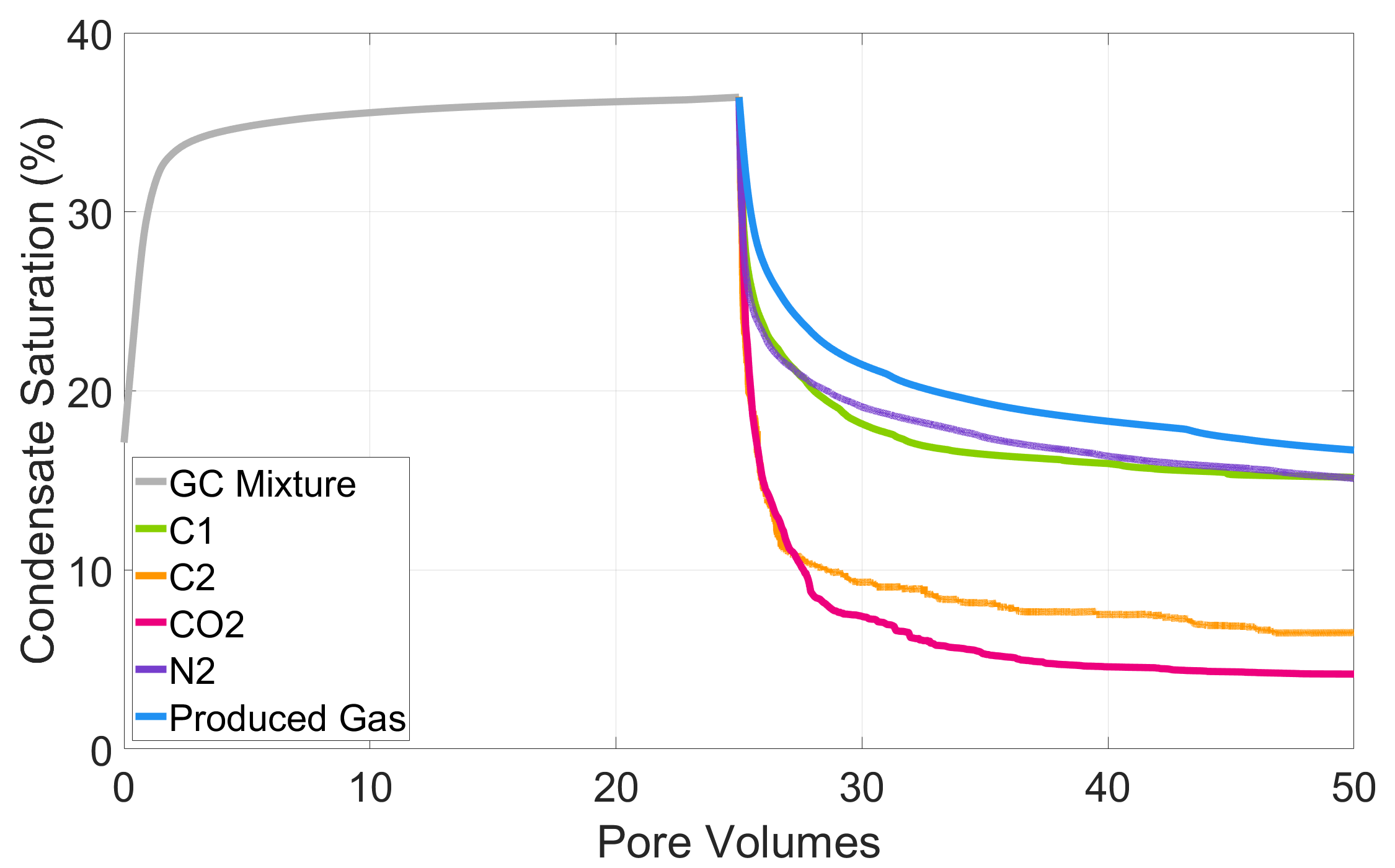}
            \caption{$P=19.5 MPa$}
         \end{subfigure}
     \end{minipage}
     \hfill
     \begin{minipage}[l]{0.45\textwidth}
         \centering
          \begin{subfigure}{1\textwidth}
            \includegraphics[width=7cm]{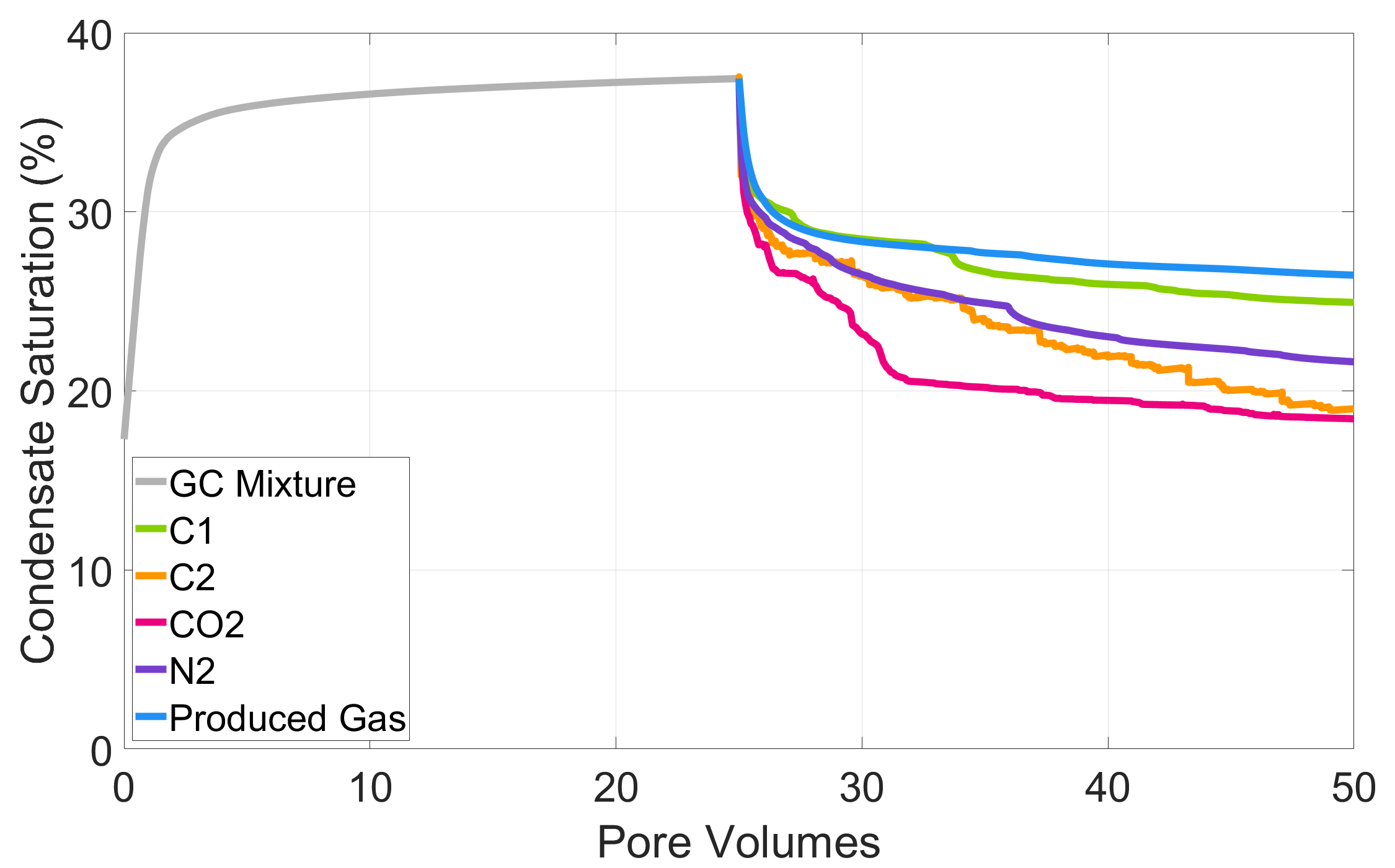}
            \caption{$P=17.75 MPa$}
         \end{subfigure}
     \end{minipage}
     \hfill{}
     \begin{minipage}[r]{0.45\textwidth}
         \centering
          \begin{subfigure}{1\textwidth}
            \includegraphics[width=7cm]{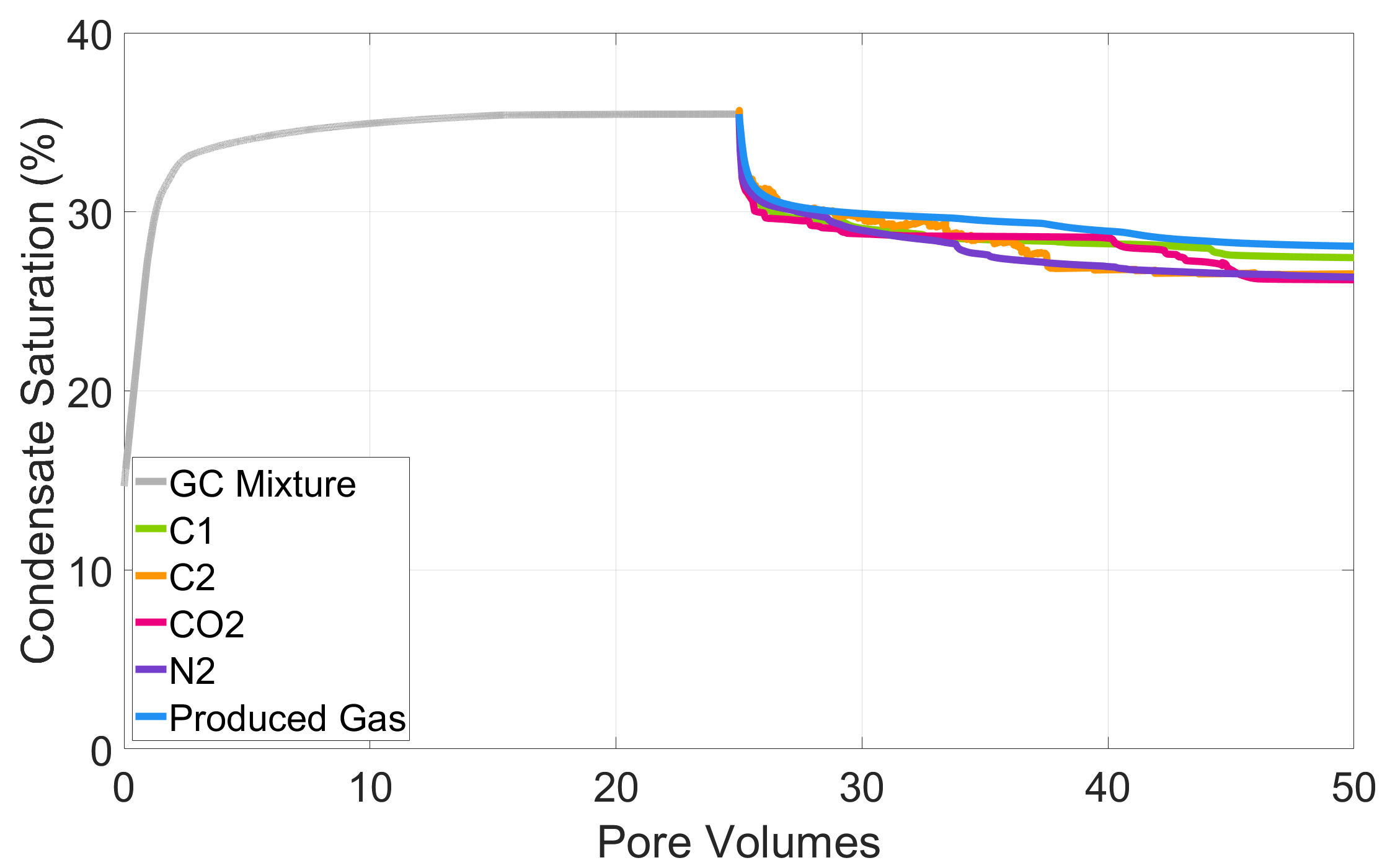}
            \caption{$P=14.75 MPa$}
         \end{subfigure}
     \end{minipage}
     \hfill
     \caption{Condensate saturation evolution with time in the networks. Injection of the reservoir mixture followed by $C_1$, $C_2$, $CO_2$, $N_2$ or produced gas.}
\label{fig:cond_sat_99}
\end{figure}

Figures \ref{fig:cond_sat_99}(a)-(f) illustrate the evolution of the condensate saturation in the networks with time, during the injection of the gas-condensate mixture followed by $C_1$, $C_2$, $CO_2$, $N_2$ or produced gas, at different pressures. The first 25 PV of injection represent the condensate buildup during the injection of the reservoir fluid, and the last 25 PV represent the condensate recovery during gas injection. At the pressure of $22 MPa$ (Fig. \ref{fig:cond_sat_99}(a)), just below the dew point pressure, $C_2$ and $CO_2$ produce almost identical effects on the accumulated condensate. Both reduced the liquid content faster and to a lower saturation than the other tested gases. $C_1$ also reduced significantly the condensate saturation, from $13.93\%$ to $1.08\%$, while $N_2$ and the produced gas displayed the lowest capacity to recover condensate, leaving $3.33\%$ and $2.74\%$ of liquid in the network, respectively. Similar results were obtained with gas injection at the pressures of $21.5$ (Fig. \ref{fig:cond_sat_99}(b)) and $21 MPa$ (Fig. \ref{fig:cond_sat_99}(c)). Even though the amount of accumulated condensate in the networks increased significantly, corresponding to $27.65\%$ at $21.5 MPa$, and $32.06\%$ at $21 MPa$, the injected gases cleared the liquid damage efficiently. This is a good indicative that, at high pressures, all tested gases could be used to support reservoir pressure and mobilize accumulated liquid. As the injection pressure is lowered below $20 MPa$, however, the ability of the gases to recover condensate is clearly reduced, as shown in Figures \ref{fig:cond_sat_99}(d) to (f). The rate of liquid re-vaporization is progressively slowed down during the injection of all tested gases, indicating that more injected volume is required for the same volume of recovered condensate, as the reservoir becomes more depleted. At the lowest tested pressure of $14.75 MPa$, all injection scenarios perform similarly, leaving most of the condensate in the porous medium after 25 PV.  These findings support the hypothesis that the timing for injection of gases in gas-condensate reservoirs is crucial for effective condensate recovery and gas flow improvement.  

\begin{figure}[H]
     \begin{minipage}[l]{0.3\textwidth}
         \centering
          \begin{subfigure}{1\textwidth}
            \includegraphics[width=4cm]{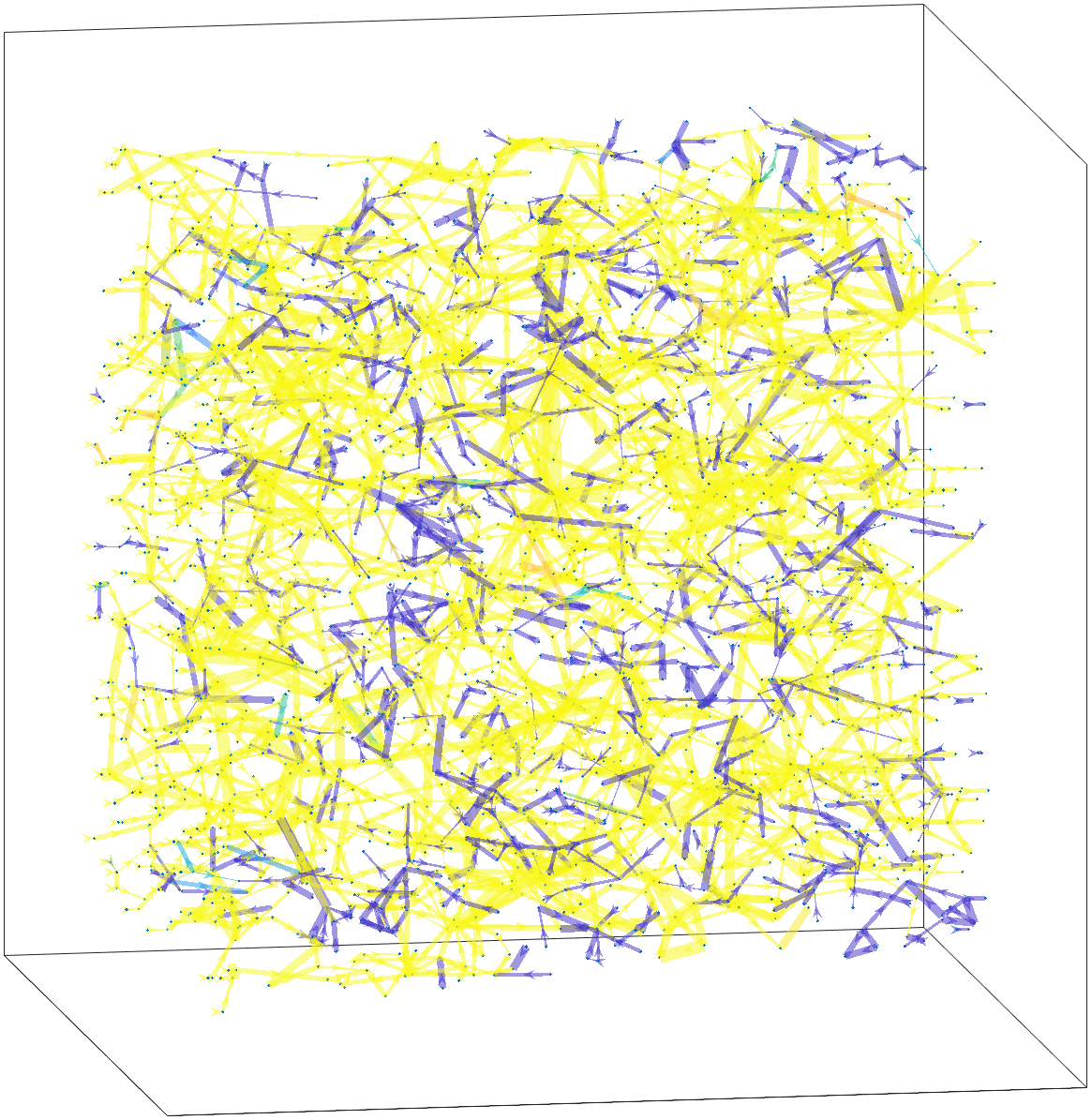}
            \caption{$P=22 MPa$}
         \end{subfigure}
     \end{minipage}
     \hfill{}
     \begin{minipage}[r]{0.3\textwidth}
         \centering
          \begin{subfigure}{1\textwidth}
            \includegraphics[width=4cm]{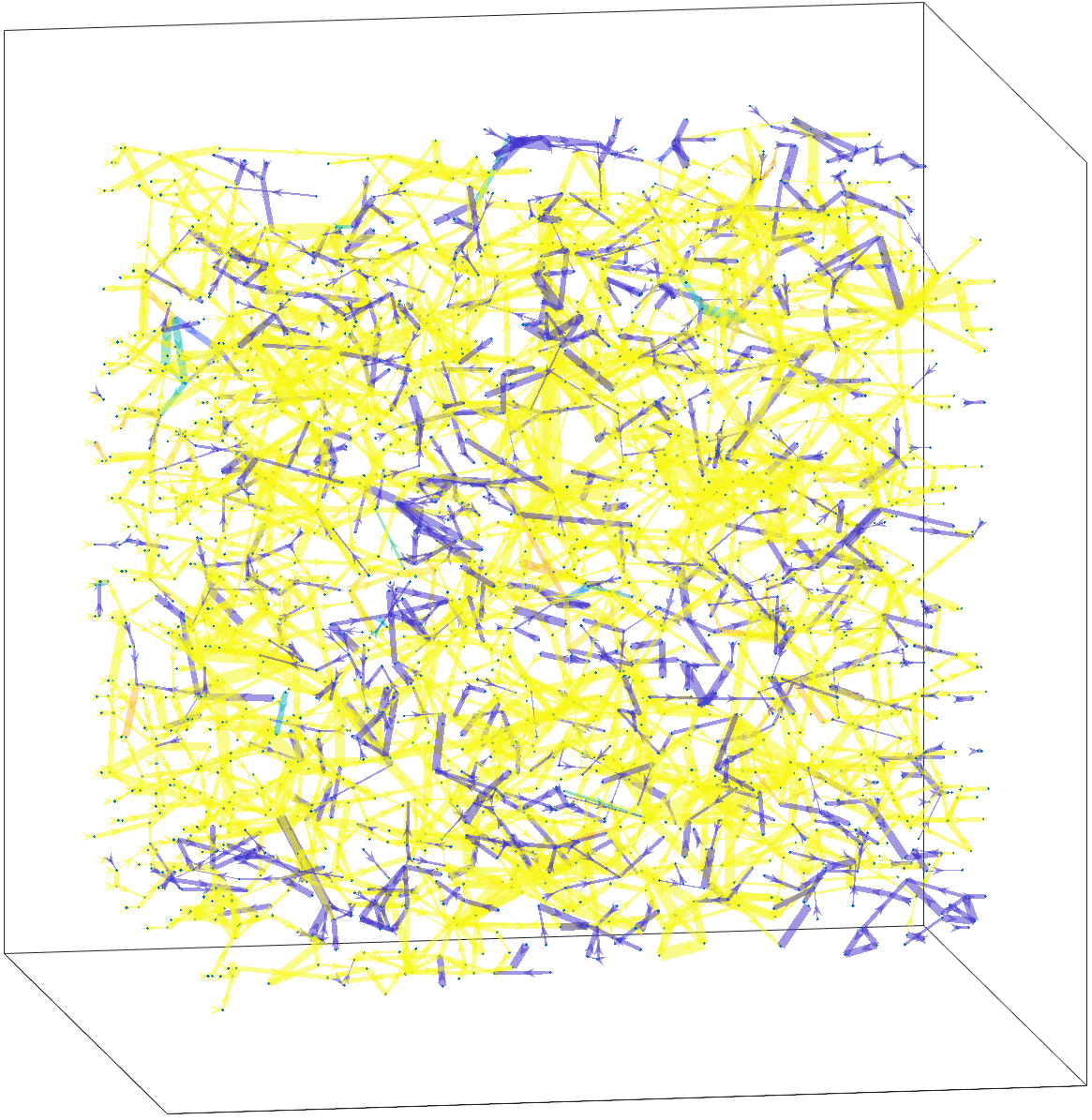}
            \caption{$P=21.5 MPa$}
         \end{subfigure}
     \end{minipage}
     \hfill
     \begin{minipage}[l]{0.3\textwidth}
         \centering
          \begin{subfigure}{1\textwidth}
            \includegraphics[width=4cm]{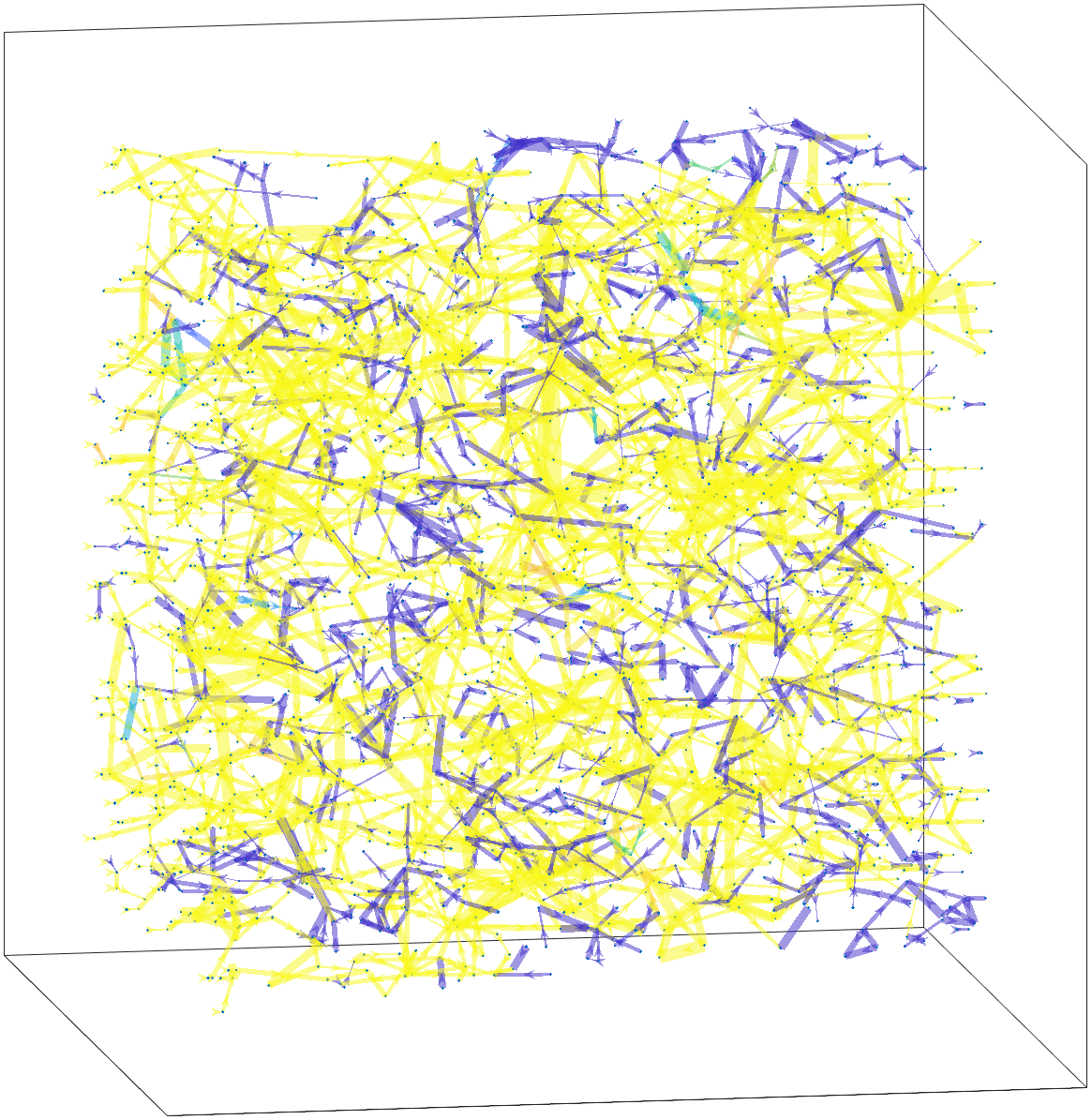}
            \caption{$P=21 MPa$}
         \end{subfigure}
     \end{minipage}
     \hfill{}
     \begin{minipage}[r]{0.3\textwidth}
         \centering
          \begin{subfigure}{1\textwidth}
            \includegraphics[width=4cm]{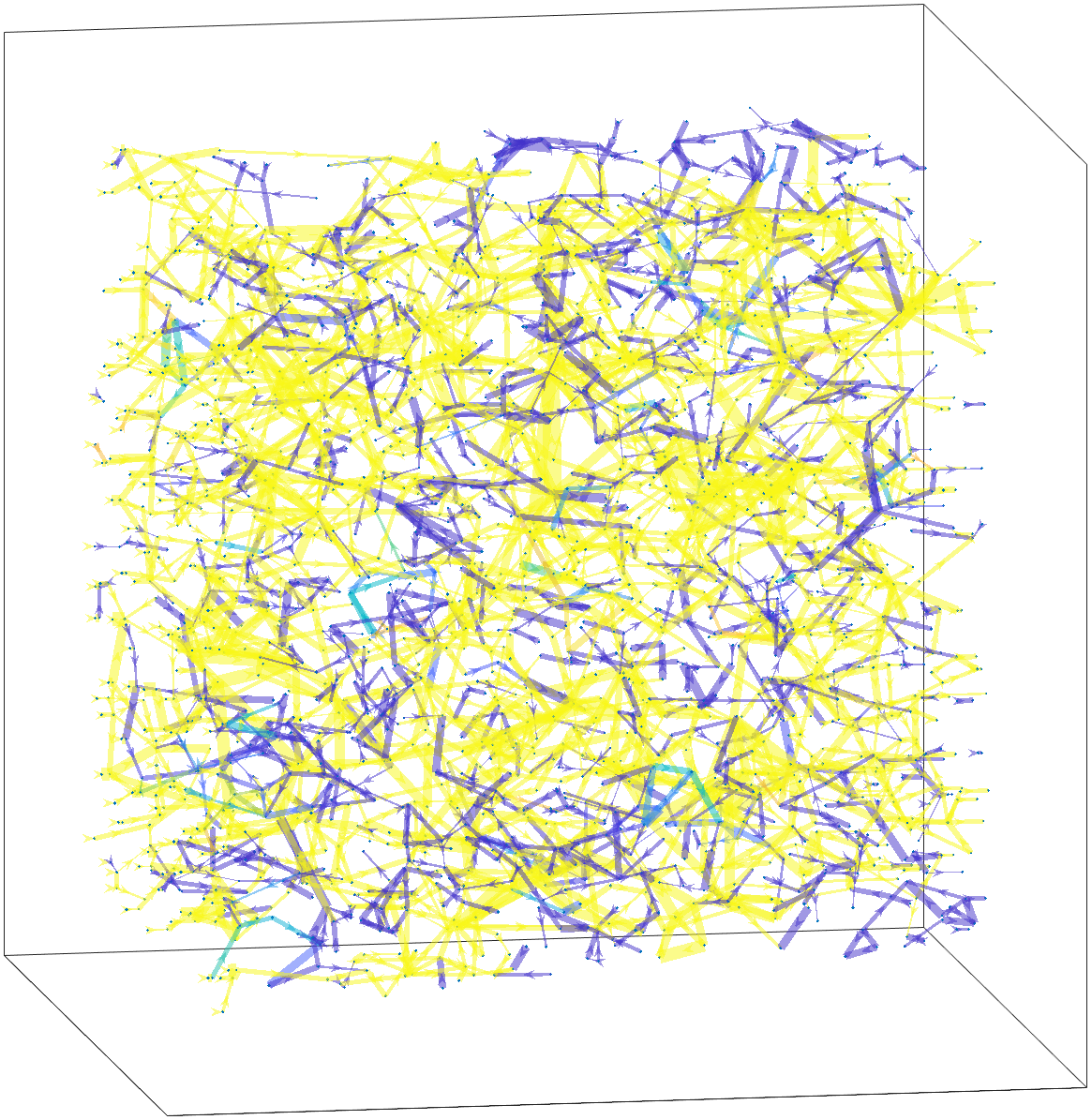}
            \caption{$P=19.5 MPa$}
         \end{subfigure}
     \end{minipage}
     \hfill
     \begin{minipage}[l]{0.3\textwidth}
         \centering
          \begin{subfigure}{1\textwidth}
            \includegraphics[width=4cm]{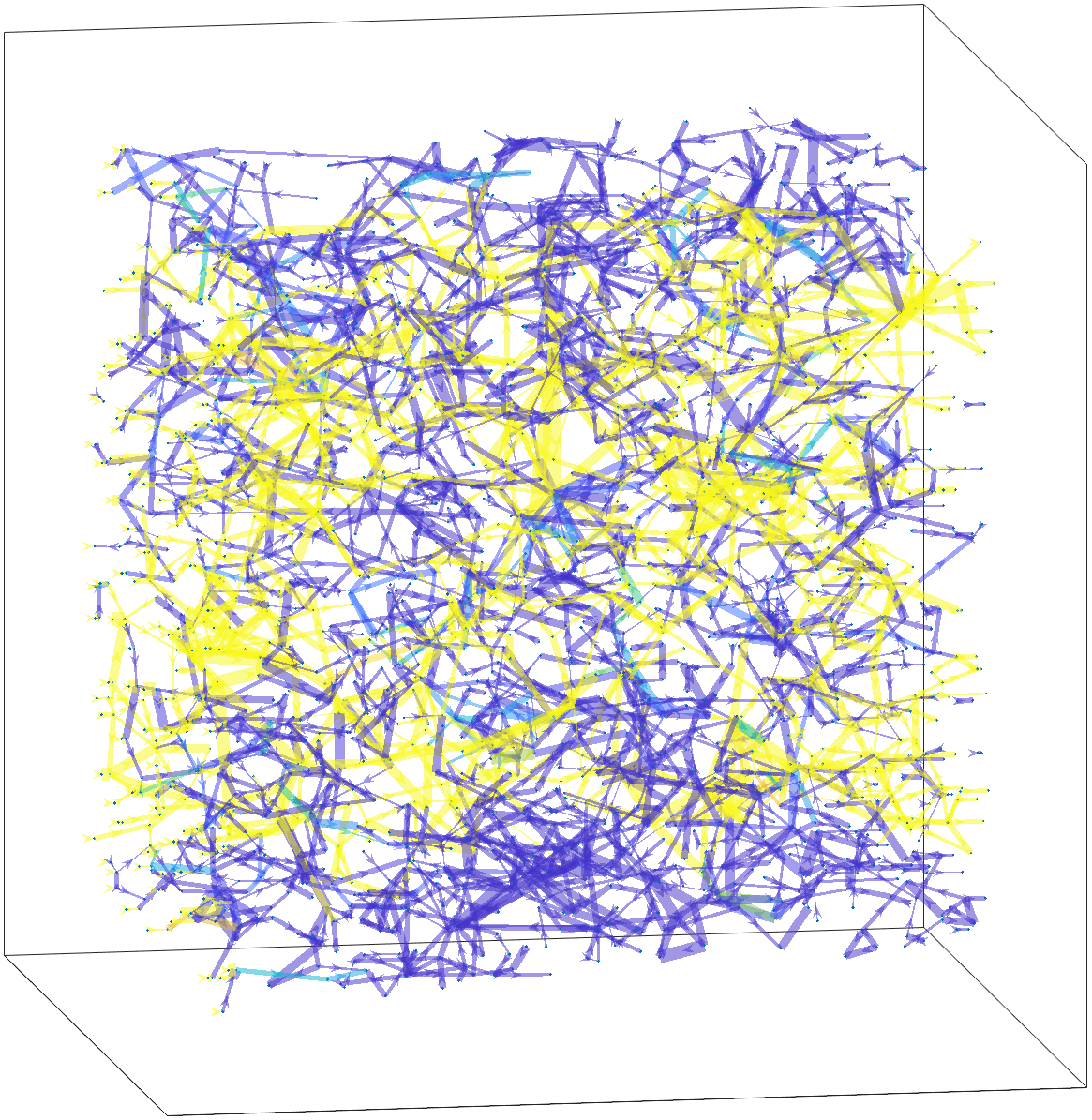}
            \caption{$P=17.75 MPa$}
         \end{subfigure}
     \end{minipage}
     \hfill{}
     \begin{minipage}[r]{0.3\textwidth}
         \centering
          \begin{subfigure}{1\textwidth}
            \includegraphics[width=4cm]{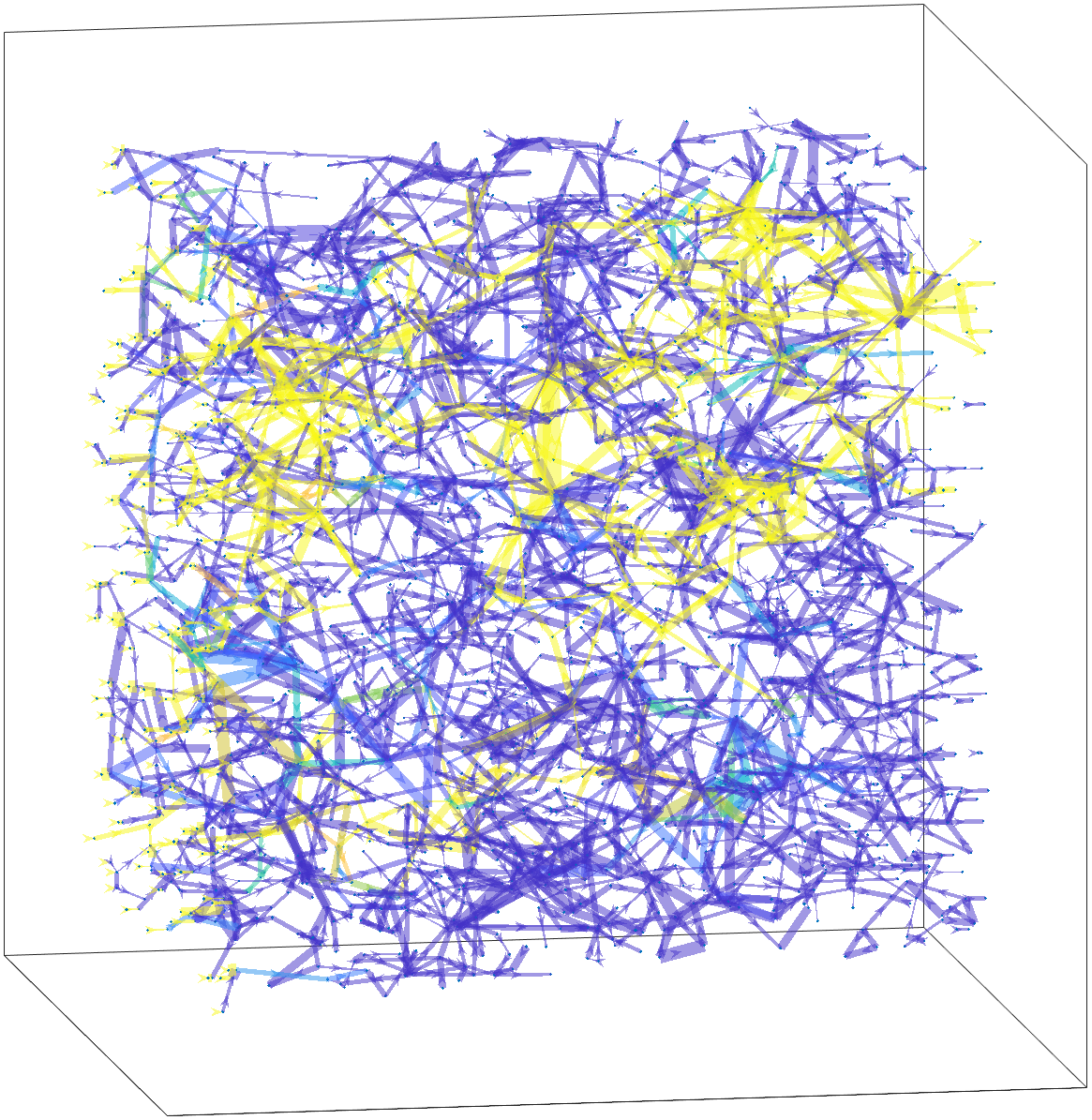}
            \caption{$P=14.75 MPa$}
         \end{subfigure}
     \end{minipage}
     \hfill
    \begin{minipage}[r]{1\textwidth}
          \begin{subfigure}{1\textwidth}
          \centering
            \includegraphics[width=10cm]{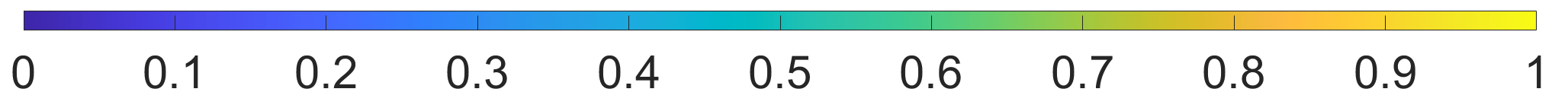}
         \end{subfigure}
     \end{minipage}
     \caption{$CO_2$ concentration in the networks after 25 PV of $CO_2$ injection at different pressures}
\label{fig:C02_sat}
\end{figure}

To further illustrate the effect of pressure in the ability of the tested gases to recover condensate, Figure \ref{fig:C02_sat} displays the $CO_2$ molar concentration in the networks following the 25 PV of $CO_2$ injection. From $22 MPa$ down to $19.5 MPa$, $CO_2$ injection's micro-sweep efficiency is gradually reduced, but most of the pore space is cleared from its liquid content. At $17.75 MPa$, the reduction in condensate recovery becomes more prominent and, at $14.75 MPa$, most of the pore space is bypassed by $CO_2$ injection.

\begin{figure}[H]
     \begin{minipage}[l]{0.45\textwidth}
         \centering
          \begin{subfigure}{1\textwidth}
            \includegraphics[width=7cm]{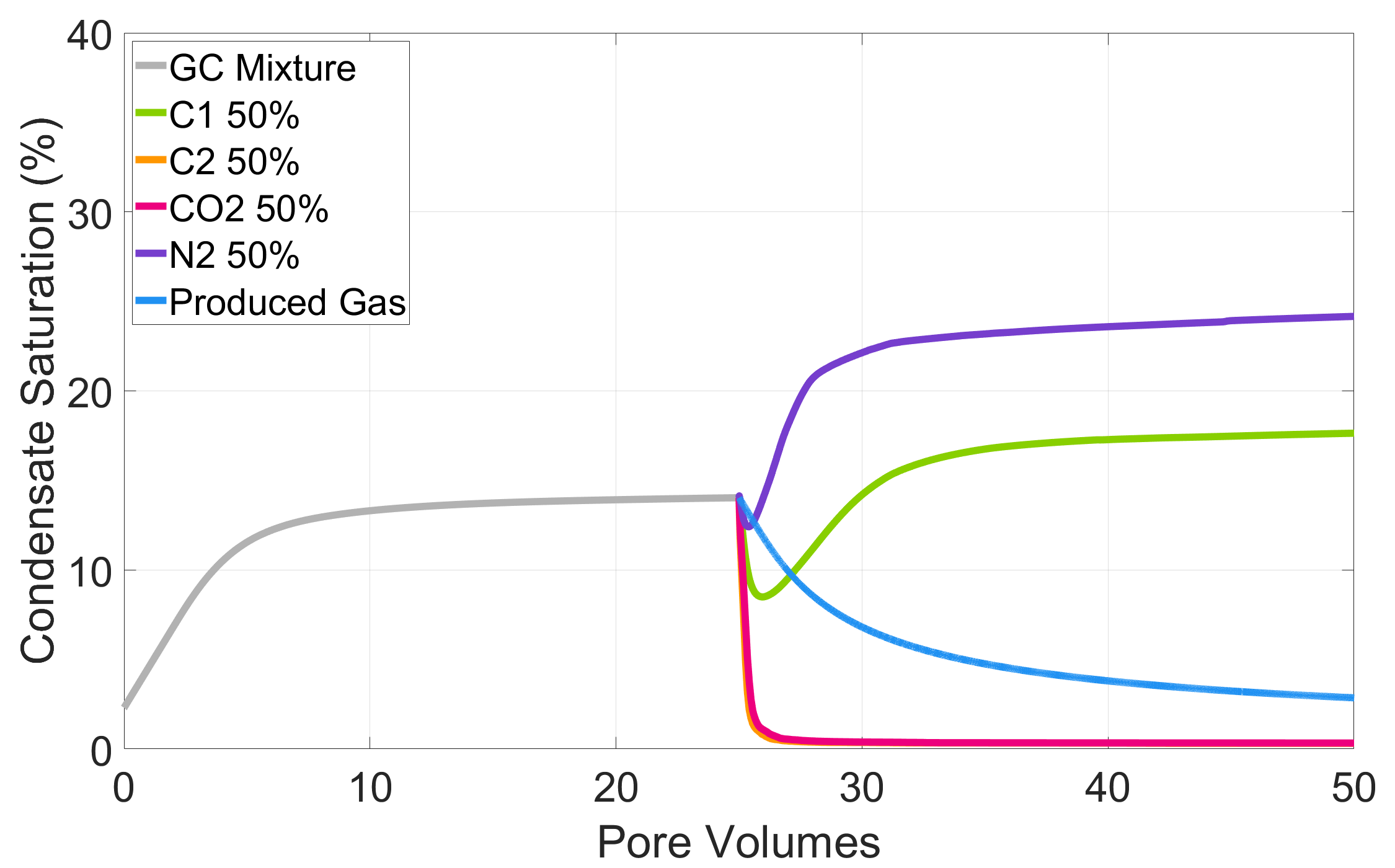}
            \caption{$P=22 MPa$}
         \end{subfigure}
     \end{minipage}
     \hfill{}
     \begin{minipage}[r]{0.45\textwidth}
         \centering
          \begin{subfigure}{1\textwidth}
            \includegraphics[width=7cm]{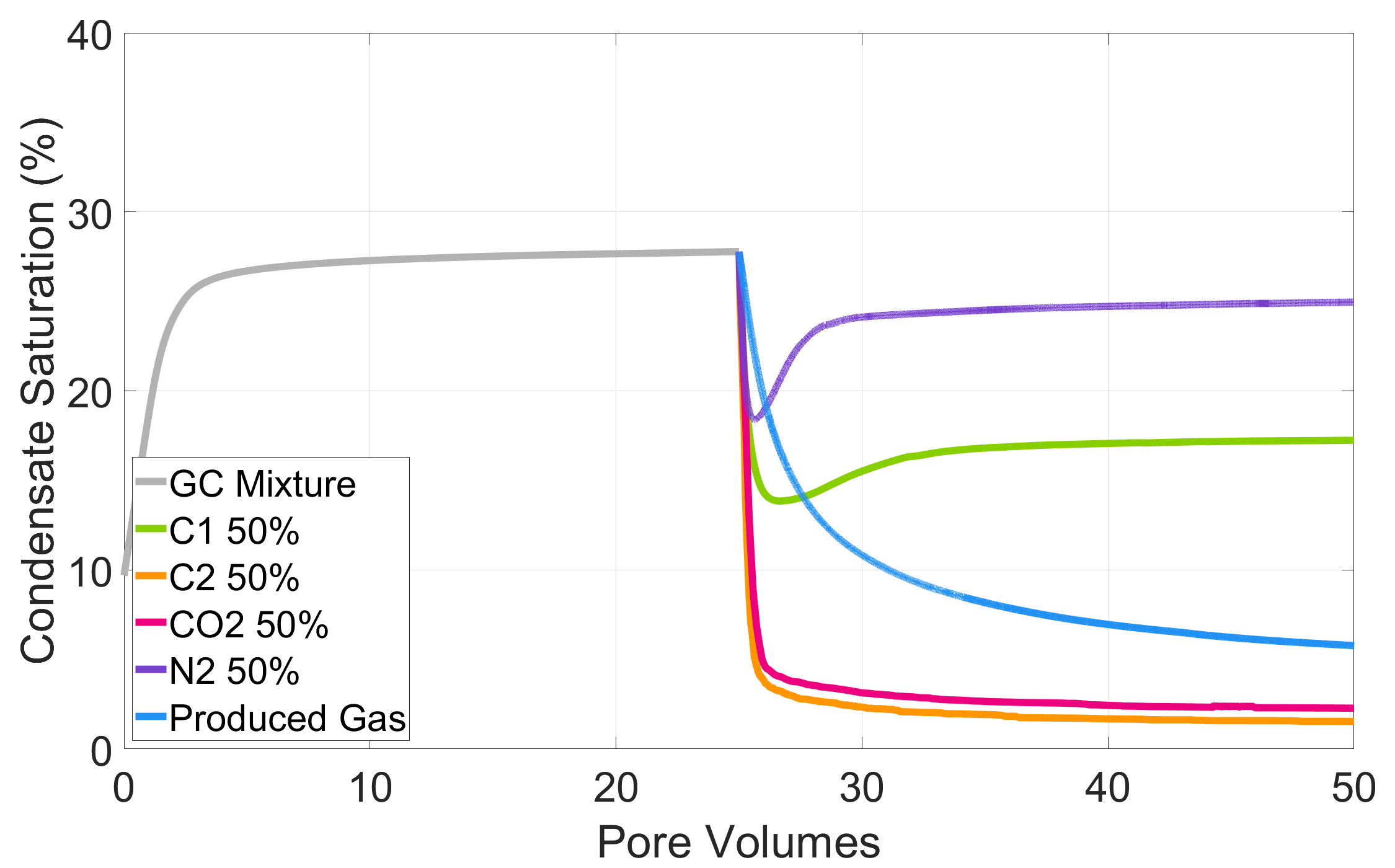}
            \caption{$P=21.5 MPa$}
         \end{subfigure}
     \end{minipage}
     \hfill
     \begin{minipage}[l]{0.45\textwidth}
         \centering
          \begin{subfigure}{1\textwidth}
            \includegraphics[width=7cm]{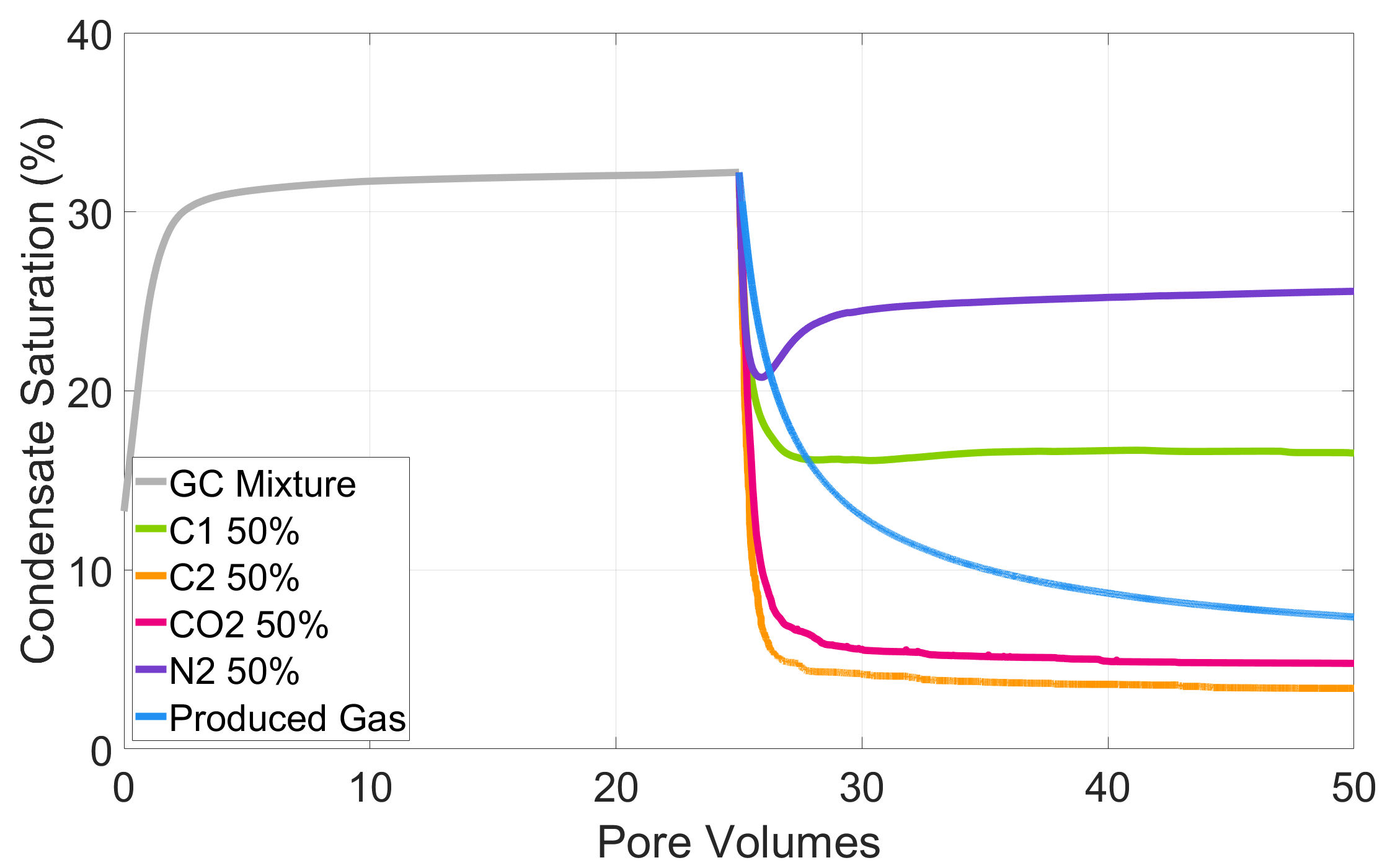}
            \caption{$P=21 MPa$}
         \end{subfigure}
     \end{minipage}
     \hfill{}
     \begin{minipage}[r]{0.45\textwidth}
         \centering
          \begin{subfigure}{1\textwidth}
            \includegraphics[width=7cm]{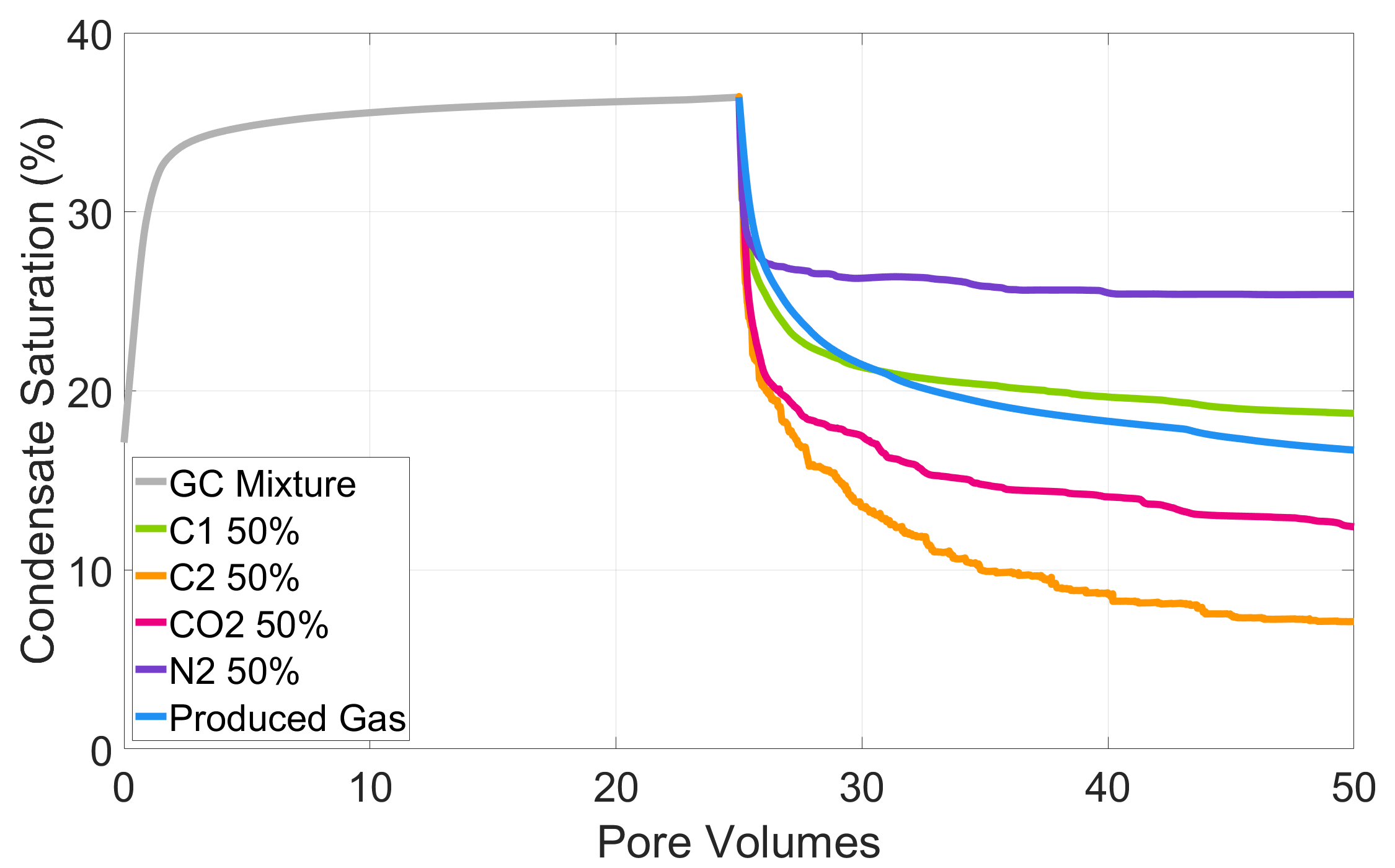}
            \caption{$P=19.5 MPa$}
         \end{subfigure}
     \end{minipage}
     \hfill
     \begin{minipage}[l]{0.45\textwidth}
         \centering
          \begin{subfigure}{1\textwidth}
            \includegraphics[width=7cm]{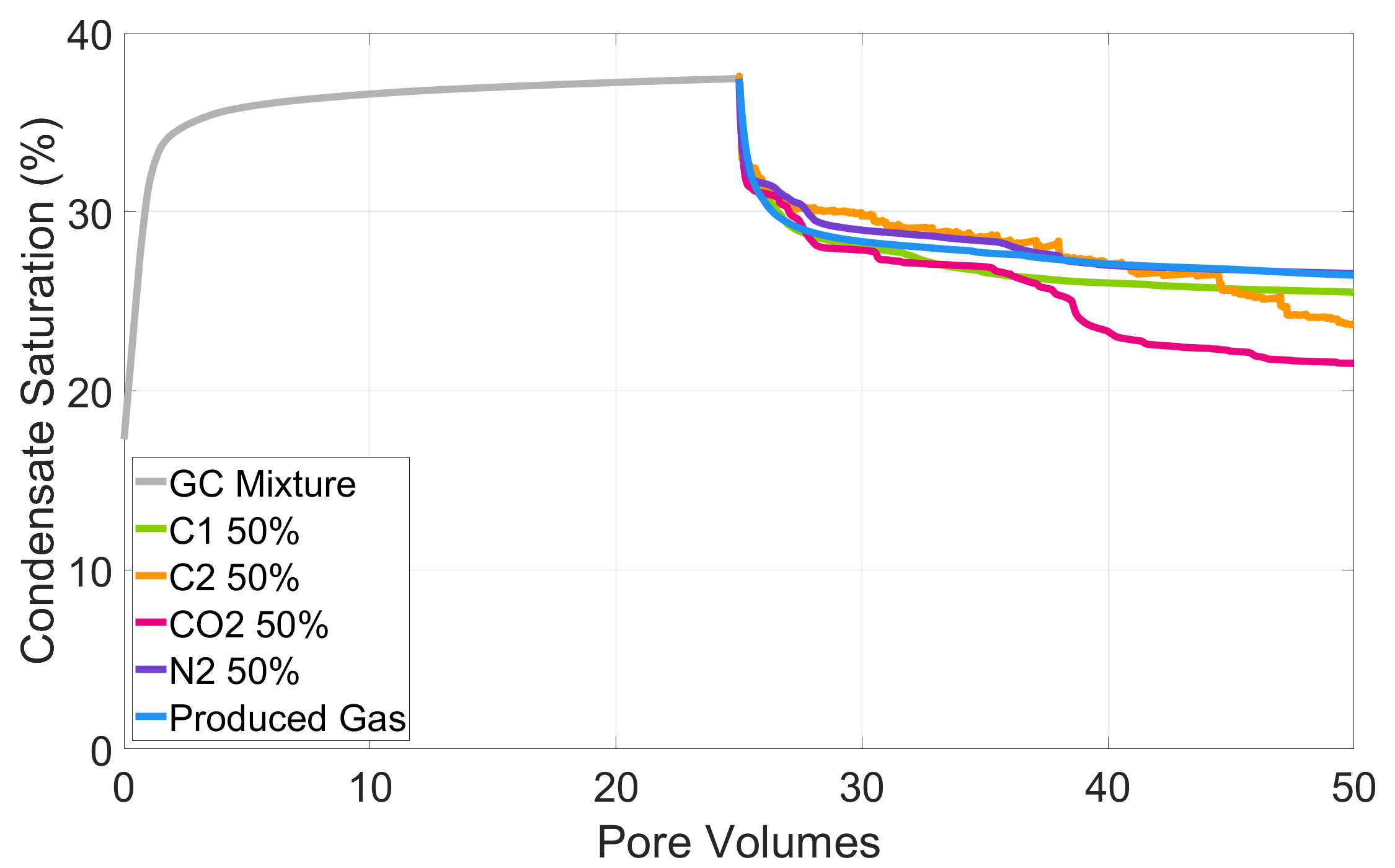}
            \caption{$P=17.75 MPa$}
         \end{subfigure}
     \end{minipage}
     \hfill{}
     \begin{minipage}[r]{0.45\textwidth}
         \centering
          \begin{subfigure}{1\textwidth}
            \includegraphics[width=7cm]{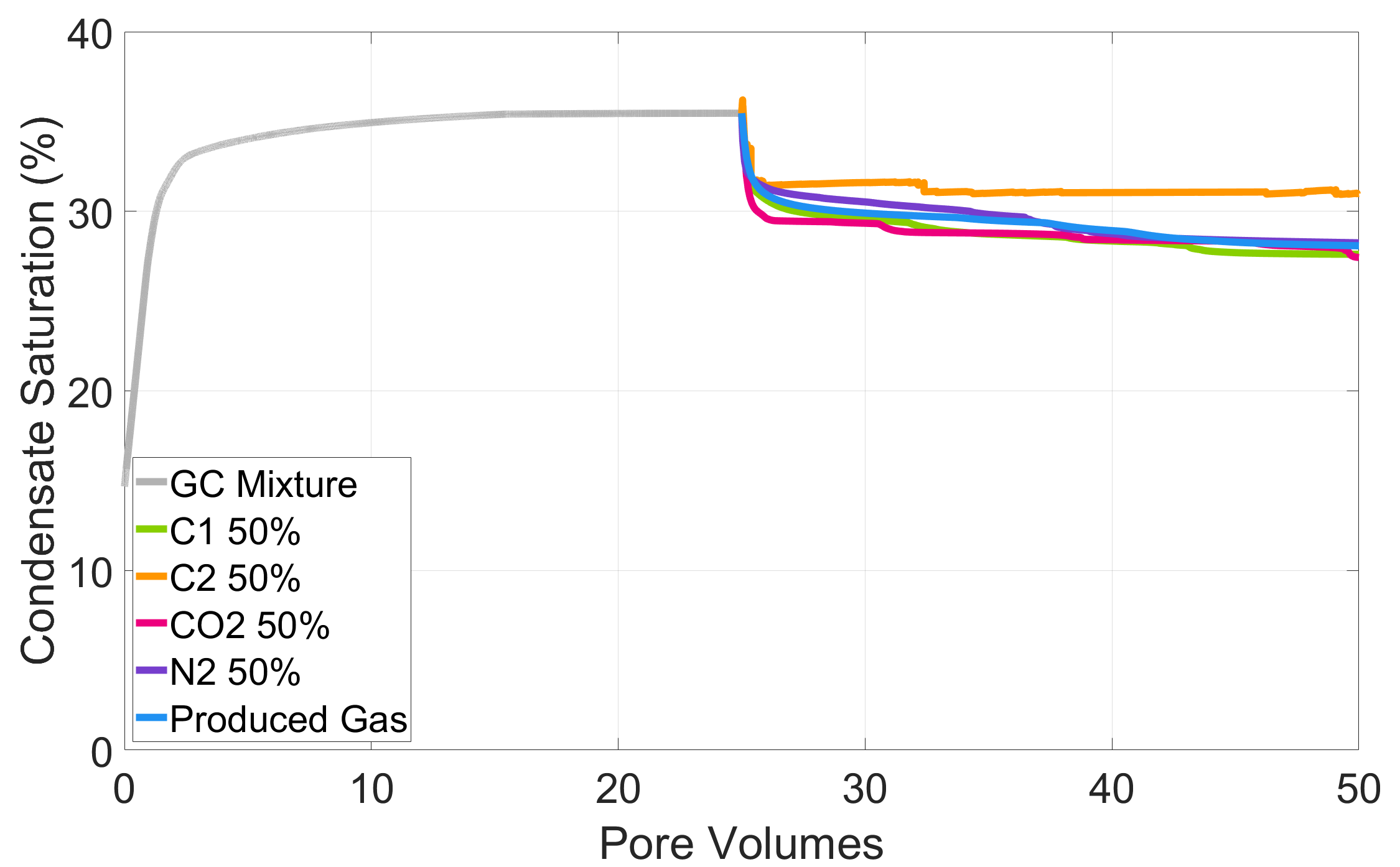}
            \caption{$P=14.75 MPa$}
         \end{subfigure}
     \end{minipage}
     \hfill
     \caption{Condensate saturation evolution with time in the networks. Injection of gases at $50\%$ molar fraction.}
\label{fig:cond_sat_50}
\end{figure}

\begin{figure}[H]
     \begin{minipage}[l]{0.45\textwidth}
         \centering
          \begin{subfigure}{1\textwidth}
            \includegraphics[width=7cm]{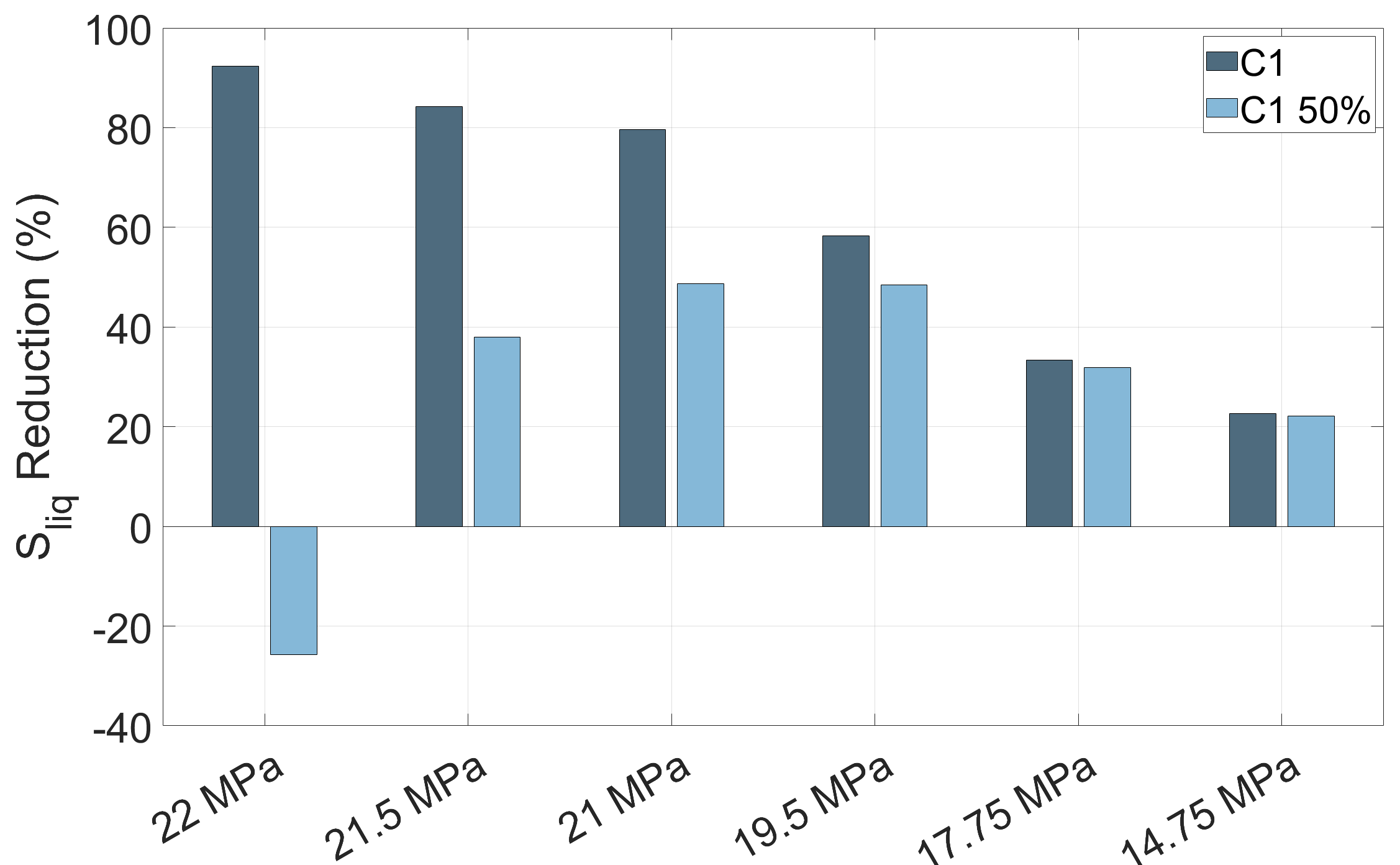}
            \caption{$C_1$ Injection}
         \end{subfigure}
     \end{minipage}
     \hfill{}
     \begin{minipage}[r]{0.45\textwidth}
         \centering
          \begin{subfigure}{1\textwidth}
            \includegraphics[width=7cm]{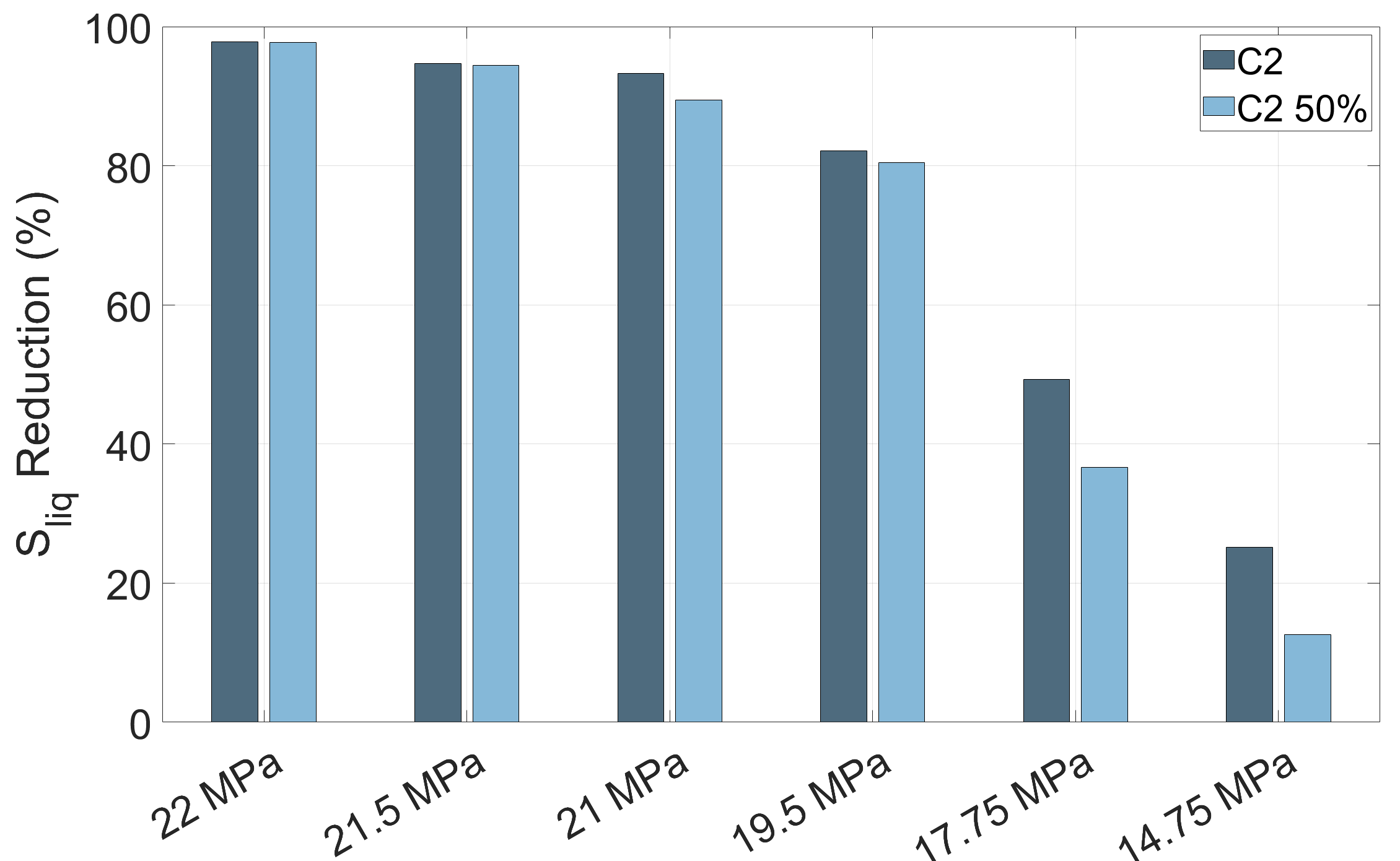}
            \caption{$C_2$ Injection}
         \end{subfigure}
     \end{minipage}
     \hfill
     \begin{minipage}[l]{0.45\textwidth}
         \centering
          \begin{subfigure}{1\textwidth}
            \includegraphics[width=7cm]{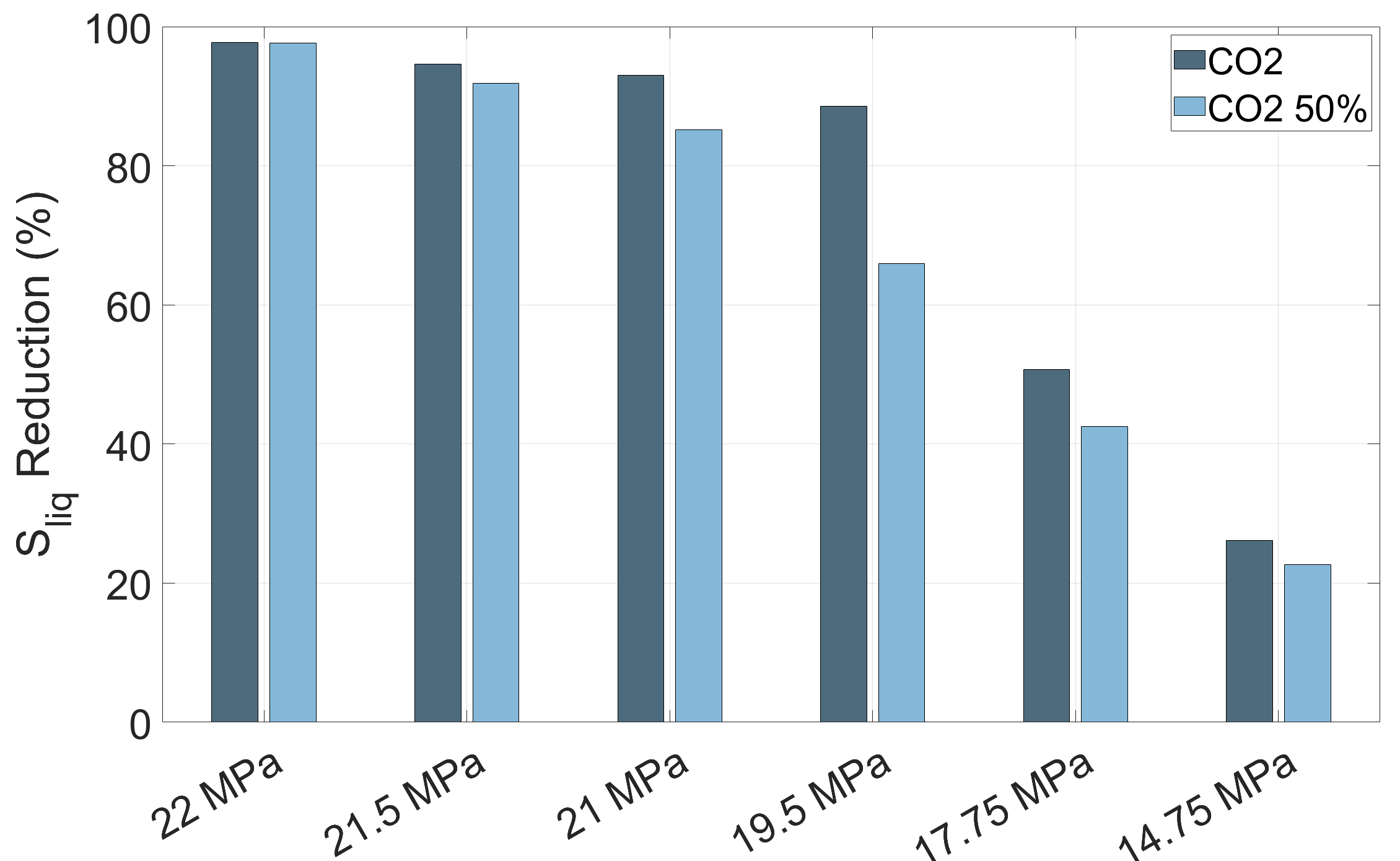}
            \caption{$CO_2$ Injection}
         \end{subfigure}
     \end{minipage}
     \hfill{}
     \begin{minipage}[r]{0.45\textwidth}
         \centering
          \begin{subfigure}{1\textwidth}
            \includegraphics[width=7cm]{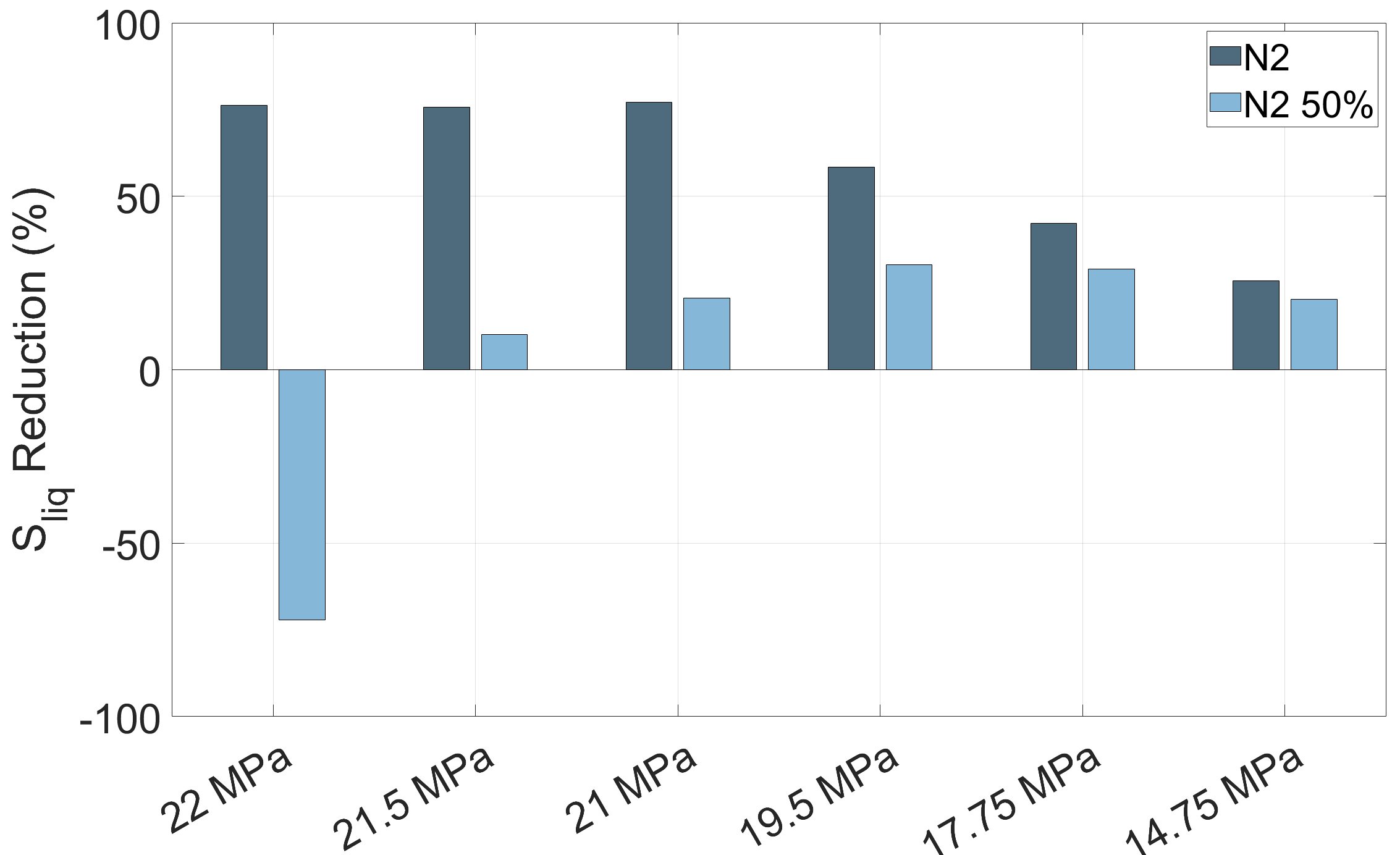}
            \caption{$N_2$ Injection}
         \end{subfigure}
     \end{minipage}
     \hfill

     \caption{Condensate saturation reductions after the injection of different gases, both pure or mixed with $50\%$ in moles with the produced gas.}
\label{fig:SL_50_99}
\end{figure}

The condensate saturation evolution during the injection of the gas mixtures containing $50\%$ in moles of the produced gas and $50\%$ of $C_1$, $C_2$, $CO_2$ or $N_2$ is presented in Figure \ref{fig:cond_sat_50}. Additionally, a comparison between the liquid saturation reductions achieved with the injection of pure $C_1$, $C_2$, $CO_2$ or $N_2$ and their mixtures with the produced gas is presented in Figure \ref{fig:SL_50_99}. 

When injected mixed with the produced gas, $C_1$ and $N_2$ were rather ineffective to reduce the accumulated condensate in the porous medium. At the highest evaluated pressure of $22 MPa$, both gases even enlarged the liquid content in the networks, as seen in Figure \ref{fig:cond_sat_50}(a). While the injection of the $C_1$ mixture led to an increase of $25.69\%$ in the liquid saturation in the network, the $N_2$ mixture enlarged it by $76.26\%$. This result is related to the negative effect of $C_1$ and $N_2$ on the reservoir mixture's liquid dropout at high pressures, as shown in Figure \ref{fig:LDO}(b). Yet, even at lower pressures, where the effects of $C_1$ and $N_2$ on condensate dropout become positive, very high liquid saturations remained in the networks after gas injection. This suggests that, to be successful as a condensate recovery method, the concentration of $C_1$ and $N_2$ flowing in porous media should be kept high, as the effects of mixing with the reservoir fluids can significantly reduce their ability to recover liquid. Contrarily, the results obtained with the injection of the $C_2$ and $CO_2$ mixtures showed very little sensitivity to the dilution with $50\%$ of the produced gas, especially at high pressures. 

\subsection{Recovery of heavy components}
\label{sec:rec_comp}

\begin{figure}[H]
\centering
\begin{subfigure}[t]{0.75\textwidth}
\includegraphics[width=100mm]{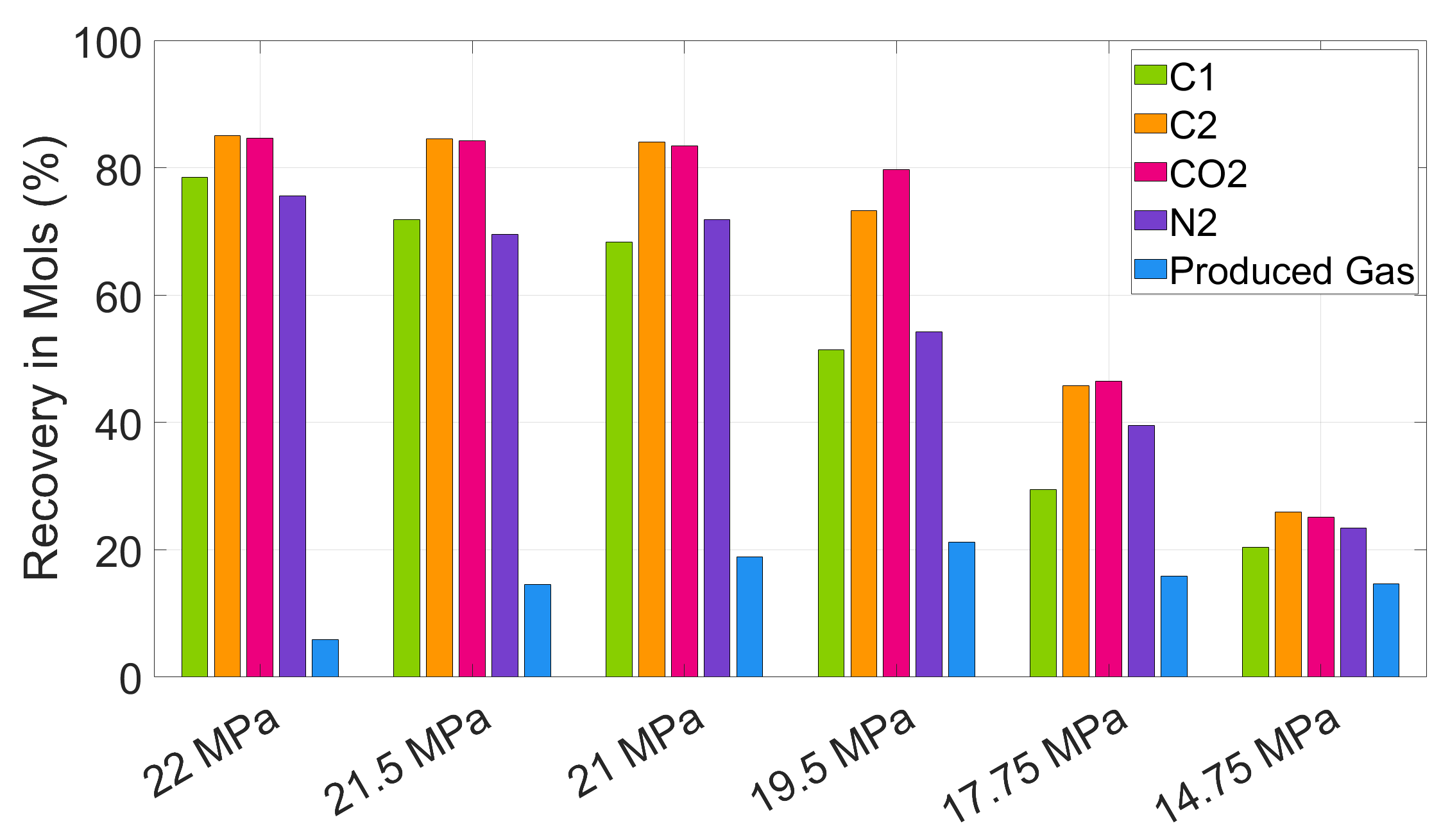}
\caption{$C_6$ Recovery}
\label{fig:C6_99}
\end{subfigure} 

\begin{subfigure}[t]{0.75\textwidth}
\includegraphics[width=100mm]{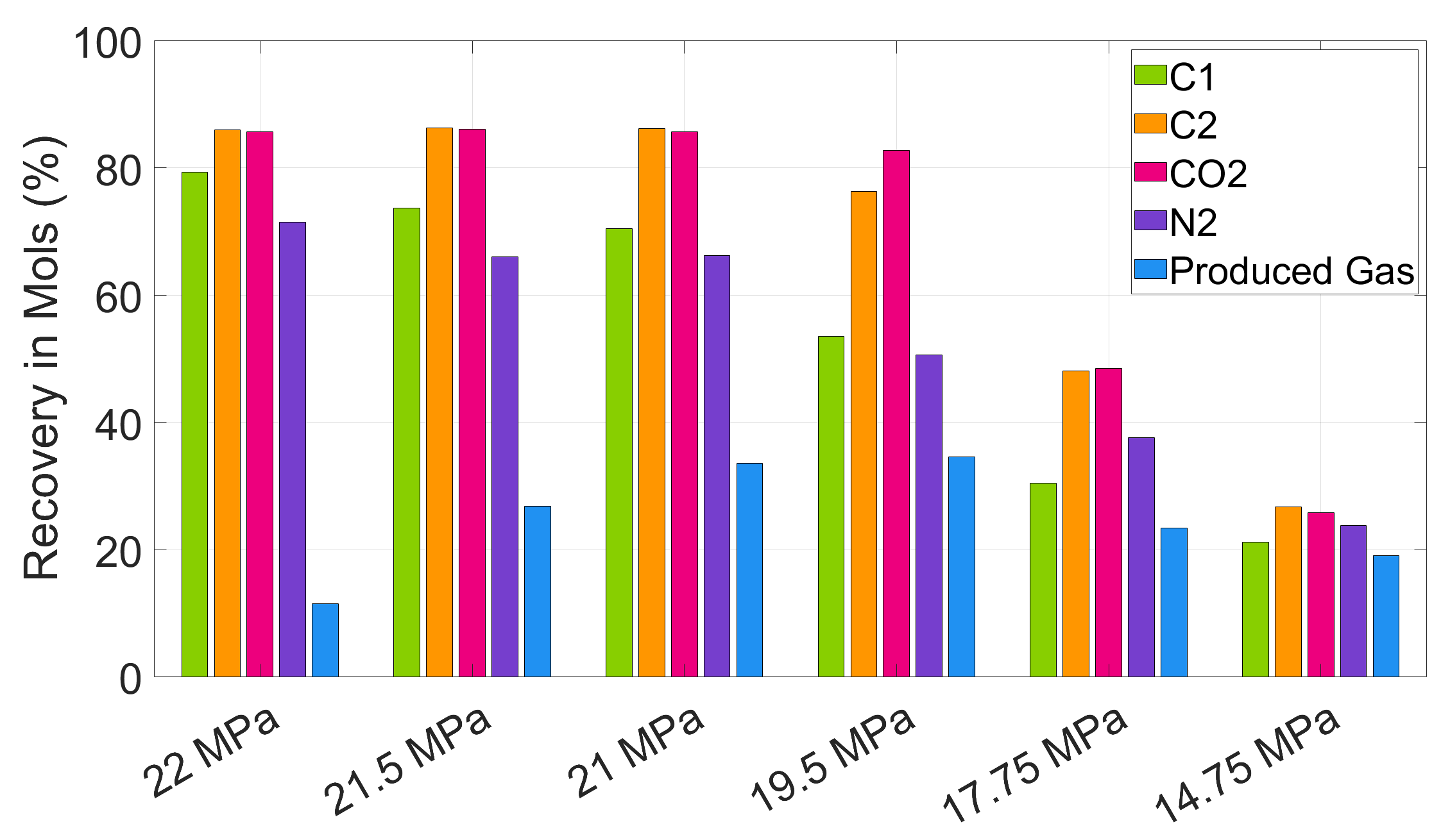}
\caption{$C_{10}$ Recovery}
\label{fig:C10_99}
\end{subfigure}

\begin{subfigure}[t]{0.75\textwidth}
\includegraphics[width=100mm]{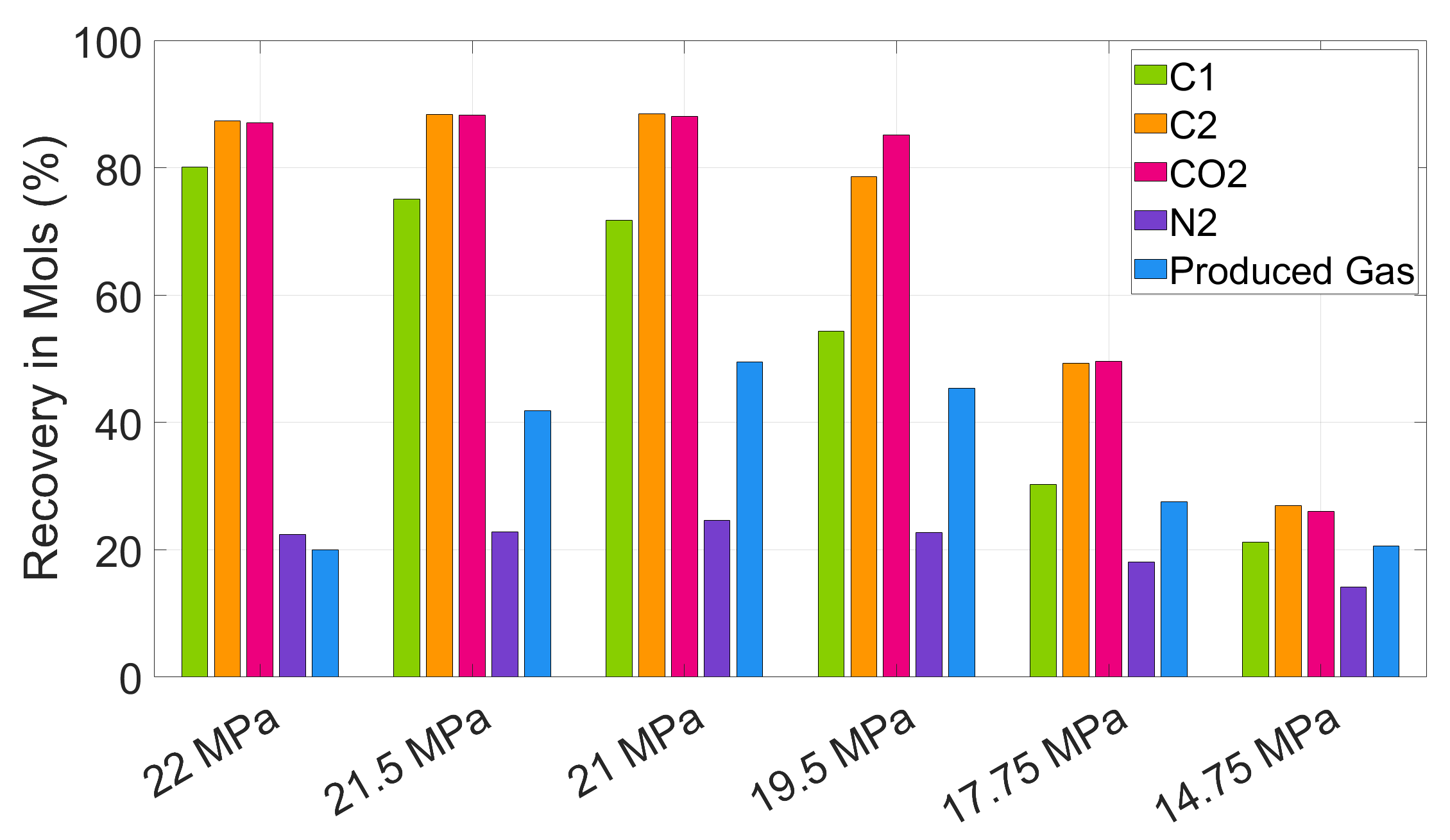}
\caption{$C_{16}$ Recovery}
\label{fig:C16_99}
\end{subfigure} 

\caption{Recovery per mixture component. Injection of gases at $99\%$ molar fraction.}
\label{fig:comp_99}
\end{figure}

\begin{figure}[H]
\centering
\begin{subfigure}[t]{0.75\textwidth}
\includegraphics[width=100mm]{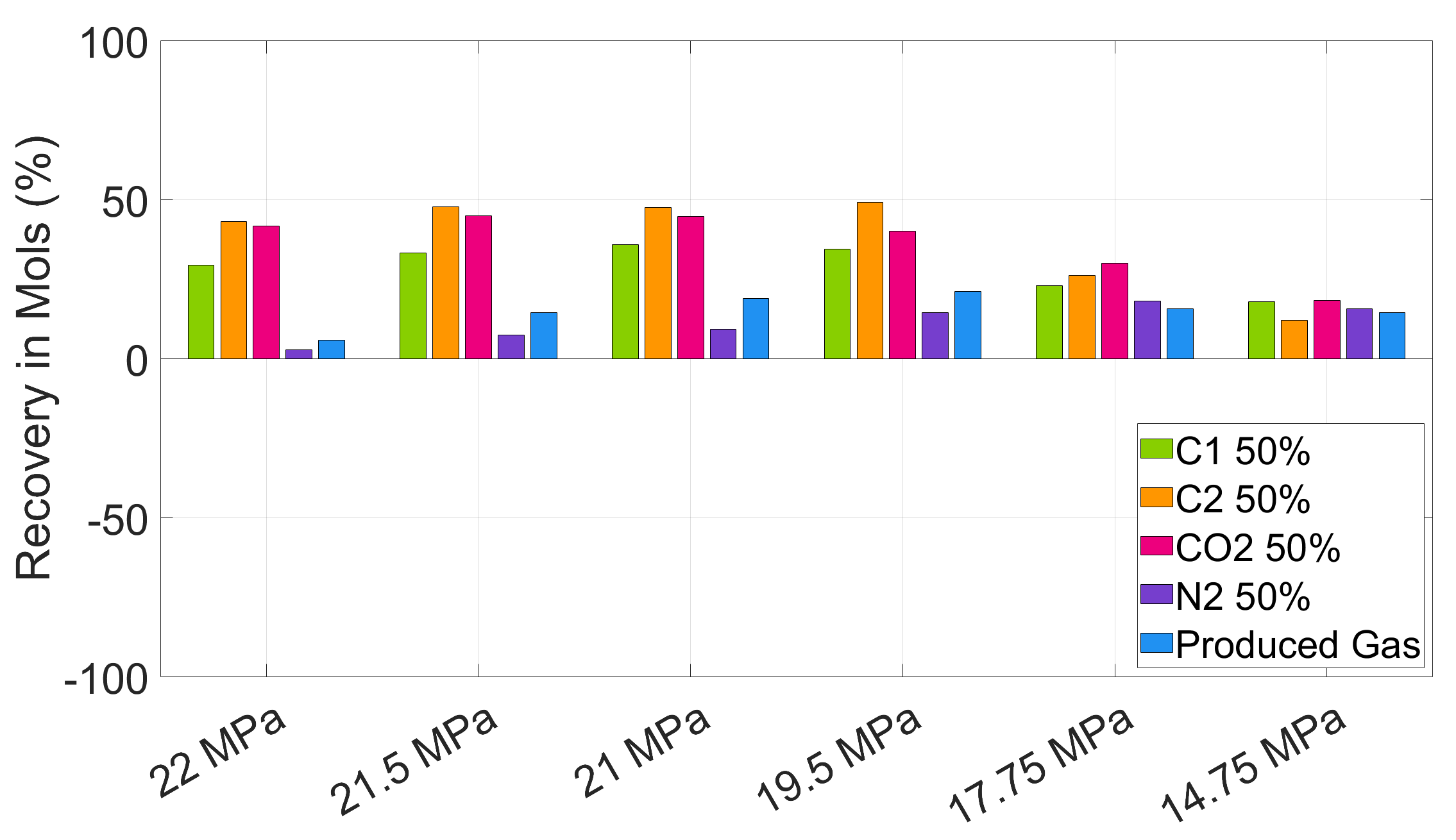}
\caption{$C_6$ Recovery}
\label{fig:C6_50}
\end{subfigure} 

\begin{subfigure}[t]{0.75\textwidth}
\includegraphics[width=100mm]{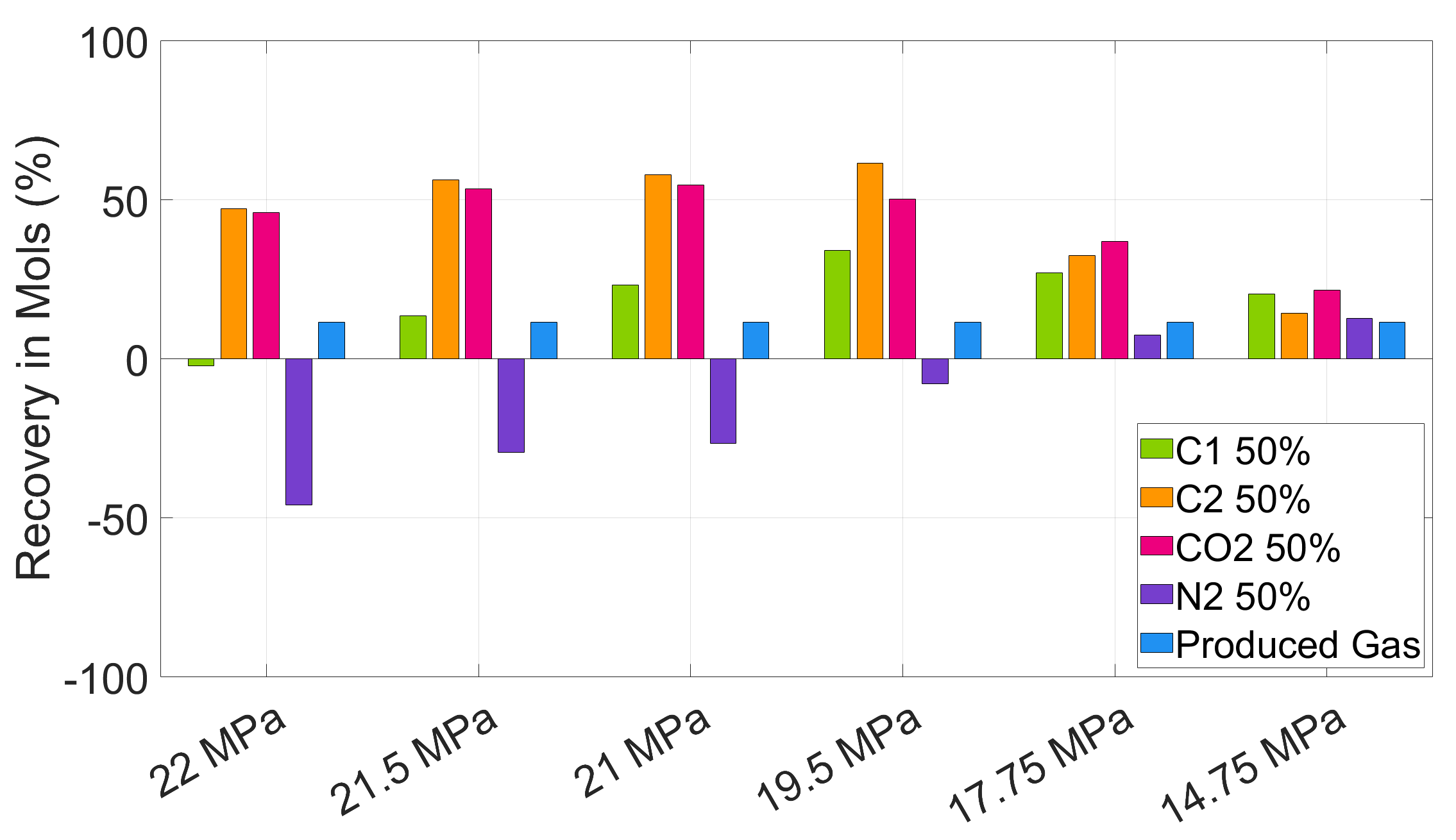}
\caption{$C_{10}$ Recovery}
\label{fig:C10_50}
\end{subfigure}

\begin{subfigure}[t]{0.75\textwidth}
\includegraphics[width=100mm]{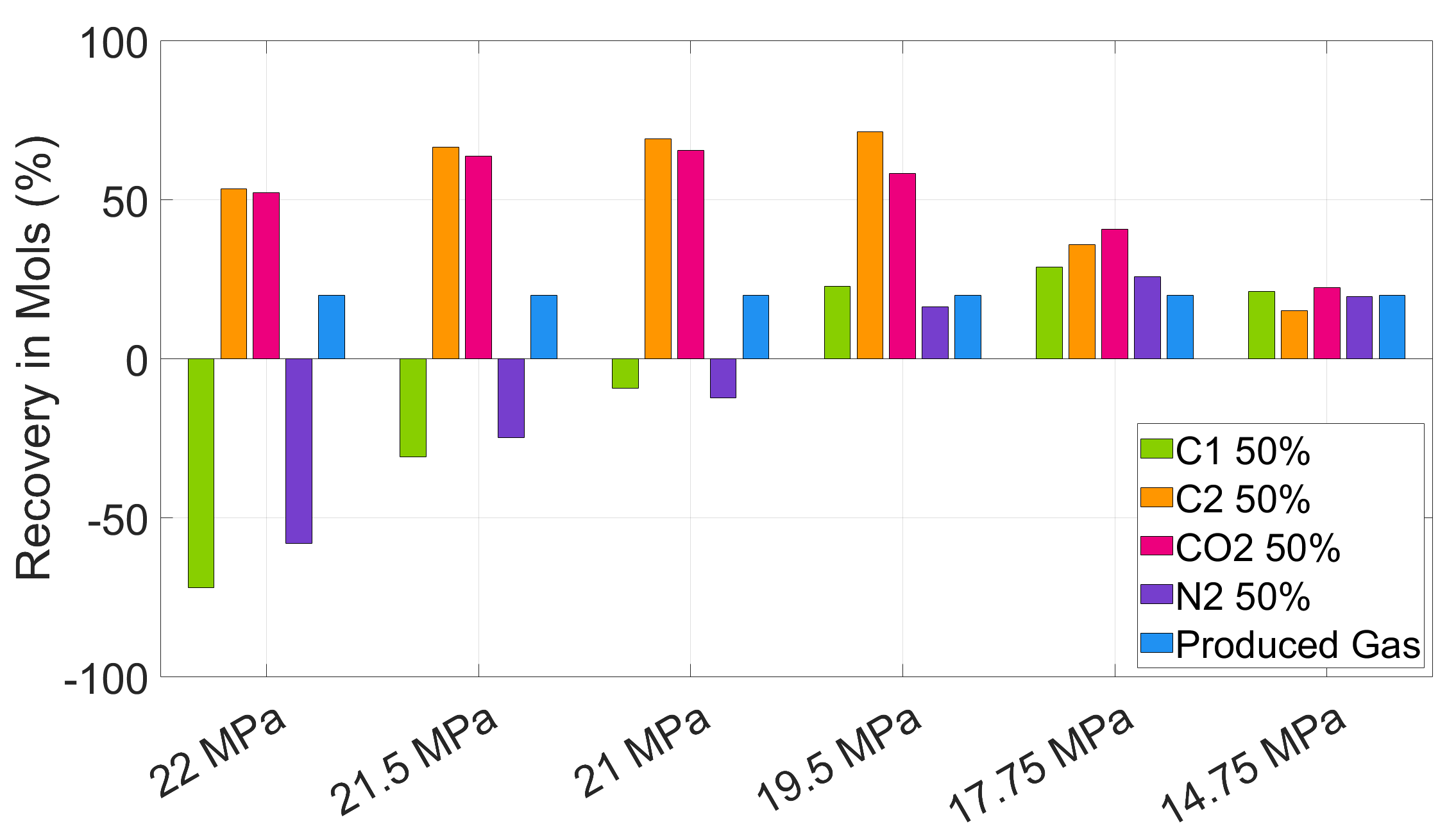}
\caption{$C_{16}$ Recovery}
\label{fig:C16_50}
\end{subfigure} 

\caption{Recovery per mixture component. Injection of gases at $50\%$ molar fraction.}
\label{fig:comp_50}
\end{figure}

 Ideally, a gas mixture injected in a gas-condensate reservoir should be able to re-vaporize not only light, but also medium and heavy components accumulated in the porous medium during condensate banking. For this reason, the recoveries of the three heaviest components of the gas-condensate fluid used in this study, namely hexane, $C_6$, decane, $C_{10}$ and hexadecane, $C_{16}$, were quantified for all tested gas injection scenarios. The recoveries obtained with the injection of $C_1$, $C_2$, $CO_2$, $N_2$ or produced gas are presented in Figure \ref{fig:comp_99}, while the recoveries obtained by injecting mixtures of these gases with the produced gas are shown in Figure \ref{fig:comp_50}. 
 
 The results in Figure \ref{fig:comp_99} suggest that, when flowing at sufficiently high concentrations, $C_1$, $C_2$ and $CO_2$ are able to recover the three heaviest components evenly. For these cases, the recoveries of each analyzed component can be directly related to the reduction in liquid saturation shown in Figure \ref{fig:SL_50_99}. During the injection of $N_2$, however, the recovery of hexadecane was considerably lower than the recoveries of hexane and decane. Therefore, in this case, during the re-vaporization of condensate, the remaining liquid in the porous medium becomes particularly rich in $C_{16}$, impeding its recovery. This indicates that, even when flowing at high concentrations, nitrogen may be unable to retrieve the heaviest components accumulated in gas-condensate reservoirs. Finally, among the tested gases, the lowest recoveries of $C_6$, $C_{10}$ and $C_{16}$ were obtained with the injection of produced gas. In this case, even at high pressures, when the liquid saturations could be significantly reduced after gas flooding, valuable heavier components remained trapped in the porous medium.
 
 Mixing $C_1$, $C_2$, $CO_2$ or $N_2$ with the produced gas at equal molar fractions before injection impacted significantly the recovery of heavy components, as shown in Figure \ref{fig:comp_50}. The injection of the mixture of $C_1$ and produced gas resulted in superior recoveries of hexane then those obtained with the injection of produced gas only, at all tested pressures. The same effect was observed for the recovery of decane, at pressures above $22 MPa$, and hexadecane, at pressures above $21 MPa$. When compared to the injection of pure $C_1$, however, the mixture of $C_1$ and produced gas led to significant lower recoveries of heavy components. Mixing $N_2$ with the produced gas led to even less favorable results. In these cases, the recoveries of heavy components were not only significantly lower than those obtained with the injection of pure $N_2$, but also, in most cases, than those achieved after the injection of produced gas only. Additionally, at high pressures, injecting the mixture of $N_2$ and produced gas led to an increased amount of trapped $C_{10}$ and $C{16}$ in the porous medium, even for the cases where the liquid saturation was reduced ($P=21.5 MPa$ and $P=21 MPa$). Therefore, in these cases, instead of re-vaporizing heavier components, the presence of $N_2$ in the injected mixture prompted a migration of $C_{10}$ and $C{16}$ from the gas to the liquid phase. 
 
 The injection of the gas mixtures rich in $C_2$ and $CO_2$ displayed overall the best results among the tested cases. With the exception of the injection at the lowest tested pressure, adding $CO_2$ and $C_2$ to the produced gas increase significantly the recovery of the heavier components accumulated in the network. This positive outcome reinforces the hypothesis arisen in section \ref{sec:cond_sat} that $C_2$ and $CO_2$ are more suitable candidates for condensate enhanced recovery than $C_1$ and $N_2$.
 
\subsection{Gas Relative Permeability}

\begin{figure}[H]
\centering

\begin{subfigure}[t]{0.75\textwidth}
\includegraphics[width=100mm]{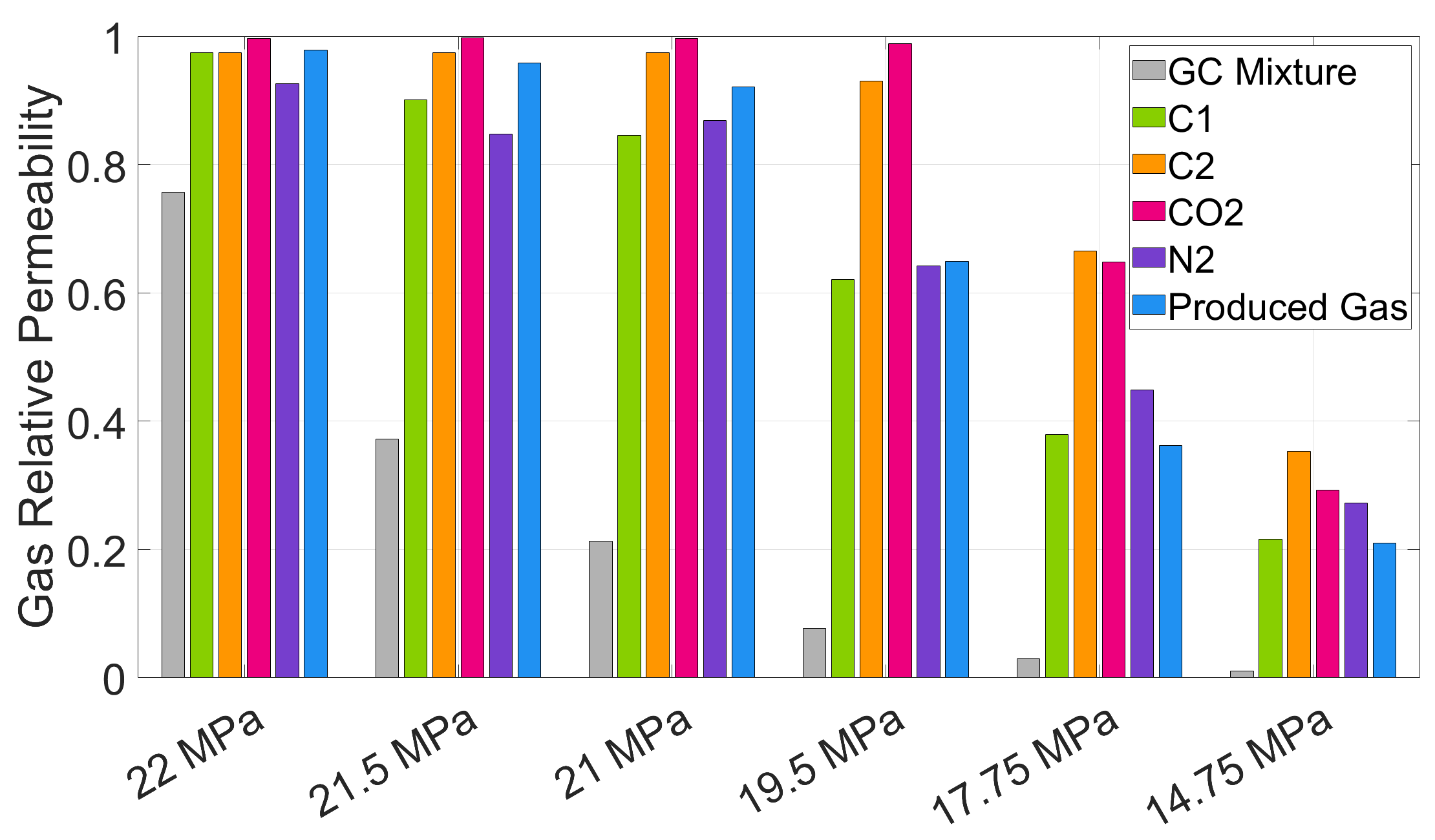}
\caption{Gas relative permeabilities before and after the injection of 25 PV of $C_1$, $C_2$, $CO_2$, $N_2$ or produced gas.}
\label{fig:krg_99}
\end{subfigure}

\begin{subfigure}[t]{0.75\textwidth}
\includegraphics[width=100mm]{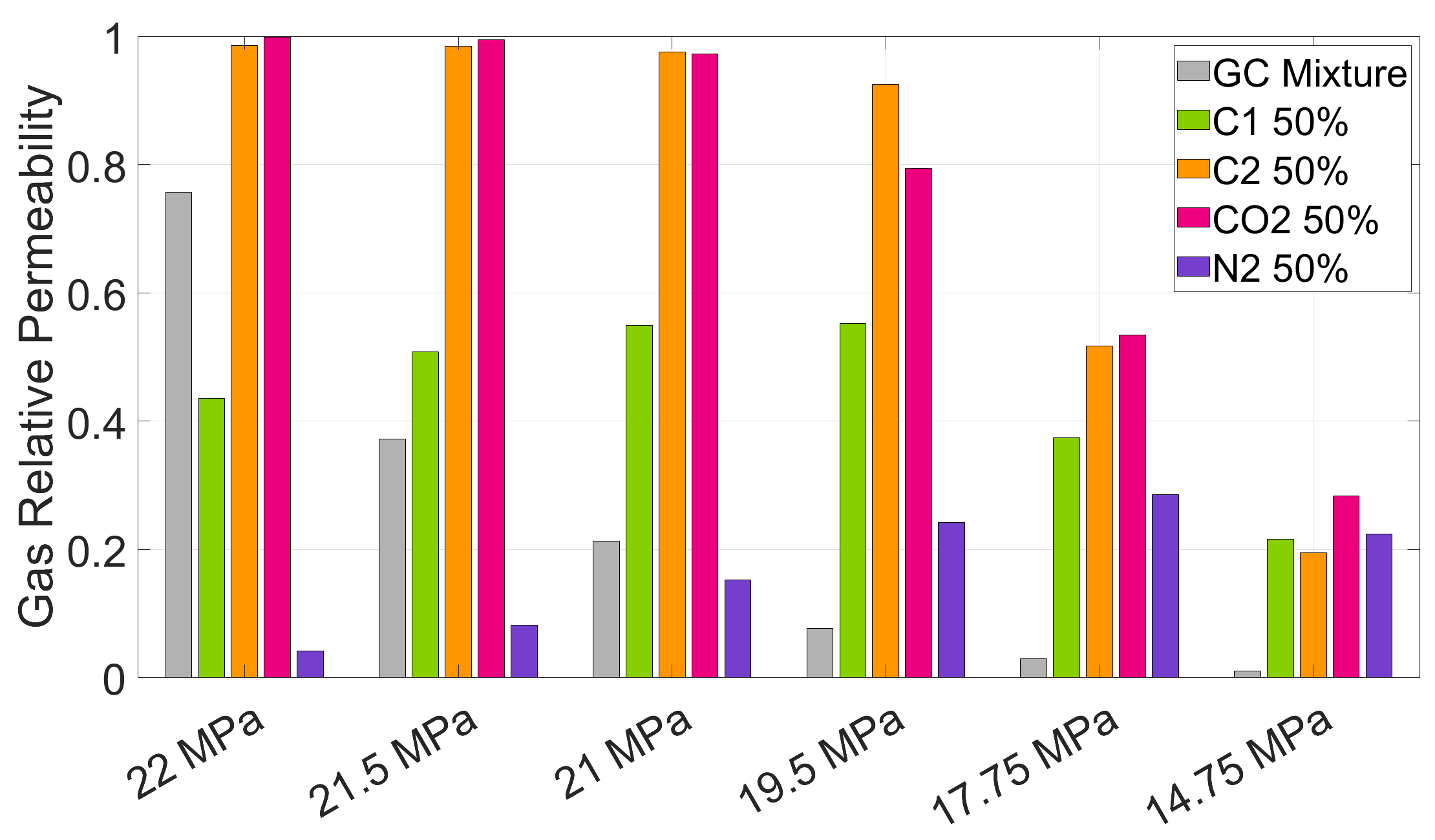}
\caption{Gas relative permeabilities before and after the injection of 25 PV of $C_1$, $C_2$, $CO_2$, $N_2$ mixed with the produced gas (50\% in moles).}
\label{fig:krg_50}
\end{subfigure} 

\caption{Gas relative permeabilities before and after gas injection.}
\label{fig:krg}
\end{figure}

As a way of directly quantifying the gas flow improvement after gas injection, in this section, we compared the gas relative permeabilities before and after the treatment. Figure \ref{fig:krg} presents the gas relative permeabilities calculated after the flow of 25 PV of the gas-condensate fluid (labeled as GC Mixture) through the network and also after the 25 PV of gas flooding, using the different tested compositions.

As can be seen in Figure \ref{fig:krg_99}, after the injection of pure $C_1$, $C_2$, $CO_2$, $N_2$ or produced gas, the gas relative permeability through the network increased significantly, at all testes pressures. At $P\geq 19.5 MPa$, injecting $C_2$ and $CO_2$ almost recovered completely the medium's damage due to condensate banking. As for the injection of $C_1$, $N_2$ and produced gas, very high values of gas relative permeability, superior to $k_{rg}=0.8$, were achieved at $P\geq 21 MPa$. Even for the two lowest tested pressures, in which cases the liquid saturation could not be  reduced to values below $20\%$, the gas relative permeabilities underwent at least a tenfold increase after gas flooding.

In Figure \ref{fig:krg_50} we can see the gas relative permeabilities before and after the injection of $C_1$, $C_2$, $CO_2$ or $N_2$ mixed with he produced gas. At high injection pressures, mixtures of $C_2$ or $CO_2$ with the produced gas could still revert successfully the damage caused by condensate accumulation, leading to $k_{rg}\approx 1$. Yet, at the same high pressures, the injection of mixtures of $C_1$ and $N_2$ with the produced gas was not as beneficial to the gas flow. At $P=22 MPa$, the $C_1$ and $N_2$ mixtures reduced the gas relative permeability from $k_{rg}=0.757$ to $k_{rg}=0.435$ and $k_{rg}=0.0417$, respectively. These negative results were expected, as the injection of these mixtures increased the liquid saturation in the porous medium at the highest tested pressure (seen in Fig.  \ref{fig:SL_50_99}b and d), due to their tendency to prompt an early condensation when mixed with the reservoir fluid (seen in Fig. \ref{fig:LDO}). At $P=21.5 MPa$ and $P=21 MPa$, however, the mixture of $N_2$ and produced gas still impacted negatively the gas flow, even though the condensate saturation in the porous medium was reduced after gas flooding. We attribute this gas flow hindrance to the effect of $N_2$ on the interfacial tension of the reservoir fluid's gas and liquid phases, explained next.

\begin{figure}[H]
    \centering
        \includegraphics[width=80mm]{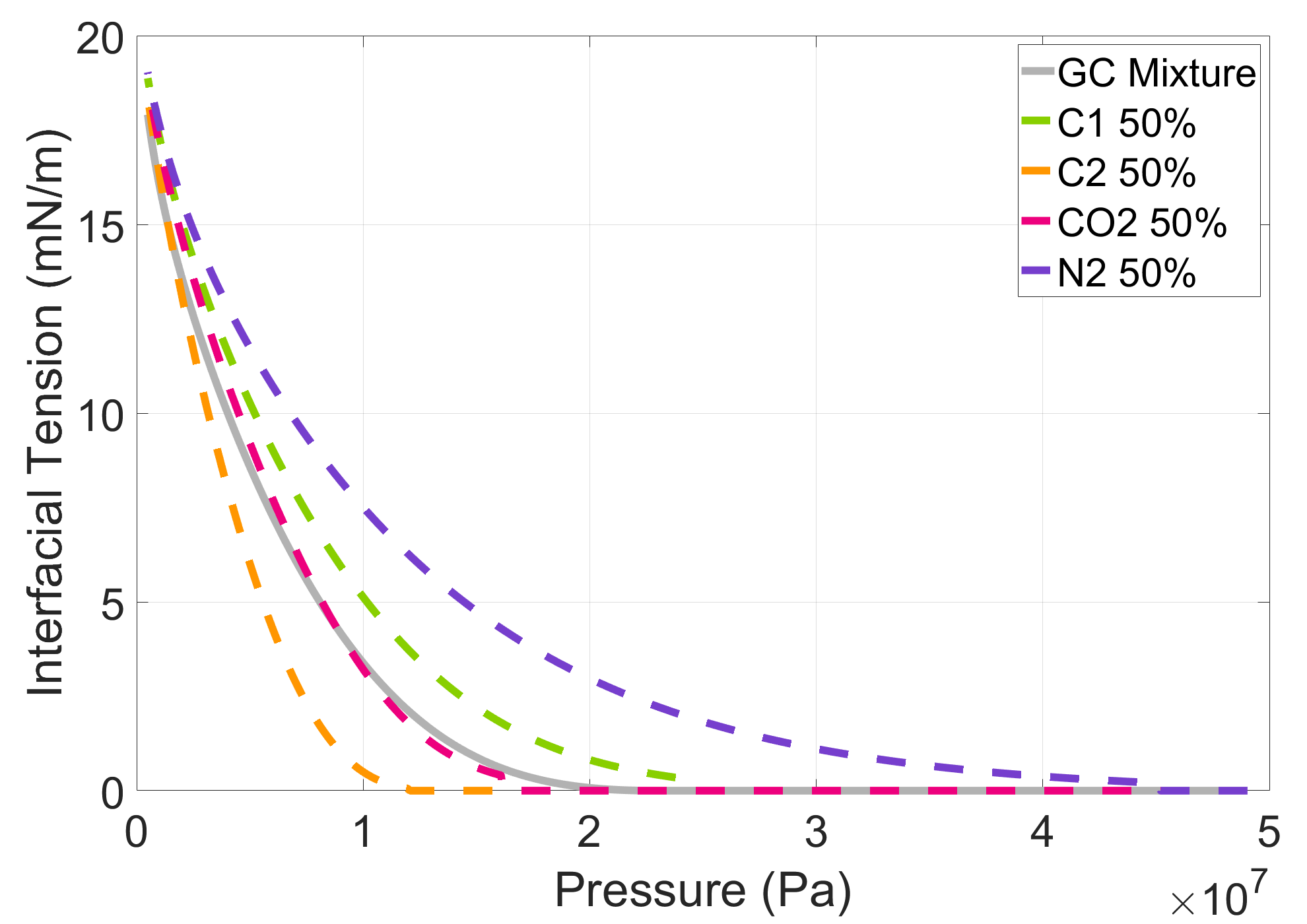}
        \caption{Calculated interfacial tension for mixtures of the reservoir fluid with $C_1$, $C_1$, $CO_2$ or $N_2$ }
    \label{fig:ift}
\end{figure}

Injecting gas on the reservoir after condensate accumulation affects not only the phases bulk properties (e.g. saturation, as seen in section \ref{sec:cond_sat}, and composition, section \ref{sec:rec_comp}),  but also interfacial properties. Considering the potential impacts on relative permeability curves \citep{al2010phase,al2011effect,al2012mobility}, these effects should be taken into account when choosing a gas mixture composition for injection in gas-condensate reservoirs. Figure \ref{fig:ift} presents the interfacial tension for the mixtures containing $50\%$ $C_1$, $C_2$, $CO_2$ and $N_2$ and $50\%$ the reservoir fluid, in moles, calculated with the correlation proposed by \citet{WK}. The results obtained with the correlation indicate that, for all evaluated pressures, mixing $C_1$ and $N2$ with the reservoir fluid leads to an increase in the interfacial tension between gas and condensate. $C_2$, on the other hand, reduces the interfacial tension, while $CO_2$ has little effect on it. 

Therefore, the significant increase in interfacial tension observed with the addition of nitrogen to the reservoir fluid mixture can strengthen the mechanism behind condensate blockage, overriding the positive effects on flow achieved with the reduction of liquid saturation. These results are particularly relevant considering that most field scale numerical studies neglect the possible effects of the injected gas composition on the micro-scale gas-condensate displacement. With this oversight, estimated recoveries obtained with $C_1$ and, especially, $N_2$ injection in gas-condensate reservoirs \citep{linderman2008feasibility,taheri2013miscible,fath2016investigation} might have been overestimated. For this reason, further studies concerning the effects of interfacial tension alteration during gas injection for condensate enhanced recovery are highly recommended.

\section{Discussion}
\label{disc}

The presented results emphasize the relevance of the injected gas composition and depletion stage in the reservoir for the success of gas injection as a condensate enhanced method. The findings suggest that $CO_2$ and $C_2$ are promising candidates for the studied method. Both gases displayed the overall highest liquid reduction potential, recovered efficiently heavy components and induced improvement in the gas-condensate relative permeability. Therefore, the use of $CO_2$ and $C_2$, either pure or mixed with another components due to cost and availability restrictions, should be further investigated for applications in partial pressure maintenance or huff-n-puff injection schemes. Pure $C_1$ injection also presented rather positive results and should be considered for injection aiming full or partial pressure maintenance. Relatively slow re-vaporization rate of condensate and potential increase in both interfacial tension and dew point pressure could, however, be a concern for the applications of $C_1$ flooding. Produced gas flooding led to similar liquid saturation reductions and gas relative permeability improvements as observed for the $C_1$ injection, but could not efficiently recover trapped heavy components from the porous medium, being considered, therefore, a slightly inferior option for condensate recovery. As for $N_2$ injection, given the feeble condensate re-vaporization tendency when mixed with the produced gas, low ability to recover heavy components and significant increase in interfacial tension, it should be carefully evaluated before being considered as a candidate for gas-condensate enhanced recovery. The relatively low injection costs \citep{linderman2008feasibility} might be canceled out by early condensation, poor two-phase flow performance and entrapment of heavy components in the porous medium. The level of depletion in the reservoir previous to the gas injection also affected notably the results. For the evaluated gas-condensate fluid, performing gas injection below the pressure associated with the maximum liquid dropout in the reservoir could not enhance significantly condensate recovery, for any tested gas composition. For this reason, the implementation of the EOR method in reservoirs at advanced stages of production might result in limited production improvement. 

The generalizability of the presented results is, however, bounded by some modeling limitations and particularities of the studied cases. First, the observed recoveries stemmed from mixing injected and reservoir fluids may depend upon the composition chosen to represent gas-condensate fluids. The extension of the analyses to different gas-condensate compositions could lead to variation in the results and, therefore, should be carried out. Additionally, porous media morphology can impact the performance of gas injection for condensate recovery, especially at low pressures, where the injected gas miscibility is limited. More heterogeneous media than the evaluated sandstone, for instance, may provide preferential paths for gas flow that deter the method's success, while very regular media could lead to increased recoveries. Finally, further studies should be performed to investigate the effects of gas injection rate on condensate recovery. As for the model limitations, the used compositional pore-network model does not take into account molecular diffusion. It has been pointed out in the literature \citep{taheri2013miscible,shtepani2006co2} that the effects of molecular diffusion during gas injection in gas-condensate reservoirs are not significant. Yet, for the analyses at low pressures, in which condensate recovery was low, implementing molecular diffusion in the model could produce more favorable results. Also, the implemented compositional formulation assumes thermodynamic equilibrium in each pore at every simulation time step. For this reason, non-equilibrium effects in the gas-condensate flow are not taken into consideration. Further investigation should be carried out on gas injection parameters, e.g. high flow rates, that may call for the incorporation of non-equilibrium effects in the model.

\section{Conclusions}

The present study provided a pore-scale evaluation of gas injection as a condensate enhanced recovery method. By using a compositional pore-network model, the effects of injected composition and pressure could be investigated by quantifying liquid saturation reduction, heavy component recovery and the impact on gas relative permeabilities. With these criteria, the performances of injecting $C_1$, $C_2$, $CO_2$, $N_2$, produced gas or their mixtures to recover the condensate accumulated in a sandstone based pore-network at different depletion stages were compared. 

The results indicated that gas injection can produce substantial flow improvement in damaged gas-condensate reservoirs, given that the injected gas composition and pressure for the treatment are adequately determined. Injecting $C_2$ and $CO_2$ produced the most favorable results among the tested gases, while $C_1$ and produced gas exhibited moderate positive effects on flow and $N_2$ displayed overall a less beneficial performance for condensate recovery.

As a preliminary investigation of gas injection in gas-condensate reservoirs at the pore-scale, this work exposed how changing the phases bulk and interfacial properties can impact significantly the two-phase flow during condensate recovery and, consequently, the method's efficiency. Interfacial tension effects, often overlooked in the field literature, showed particularly relevance in gas flow performance after gas injection. By neglecting the potential negative effects of increase in interfacial tension during the injection of $C_1$ and $N_2$, condensate recovery predictions obtained with reservoir-scale models may be overestimated. As future work, a systematic pore-scale investigation of the parameters affecting condensate recovery with gas injection, as flow rate, gas-condensate fluid composition and porous media heterogeneity should be carried out. With that, data for up-scaling pore-scale effects pertaining to particular gas injection scenarios could be generated for reservoir-scale modeling. This could lead to more realistic recovery estimations and benefit gas-condensate fields development planning.



\bibliographystyle{elsarticle-num-names} 
\bibliography{Biblio}

\end{document}